\documentclass[final, 1p, times, authoryear]{elsarticle}

\usepackage{amssymb}
\usepackage{amsthm}
\usepackage[colorlinks=true, allcolors=blue]{hyperref}     
\usepackage{bm}
\usepackage[ruled,vlined]{algorithm2e}
\usepackage{algorithmic}
\usepackage{amsmath}
\usepackage{array}
\usepackage{multirow}

\journal{Annals of Nuclear Energy}

\begin{document}

\let\today\relax
\makeatletter
\def\ps@pprintTitle{%
    \let\@oddhead\@empty
    \let\@evenhead\@empty
    \def\@oddfoot{\footnotesize\itshape
         { } \hfill\today}%
    \let\@evenfoot\@oddfoot
    }



\begin{frontmatter}
\title{Inverse Uncertainty Quantification by Hierarchical Bayesian Modeling and Application in Nuclear System Thermal-Hydraulics Codes}



\author{Chen Wang\corref{cor1}\fnref{label1}}
\cortext[cor1]{Corresponding author. Email address: chenw3@illinois.edu}
\author[label2]{Xu Wu}
\author[label1]{Tomasz Kozlowski}

\affiliation[label1]{organization={Department of Nuclear, Plasma and Radiological Engineering, University of Illinois Urbana Champaign}}
\affiliation[label2]{organization={Department of Nuclear Engineering, North Carolina State University}}

\begin{abstract}

In nuclear Thermal Hydraulics (TH) system codes, a significant source of input uncertainty comes from the Physical Model Parameters (PMPs), and accurate uncertainty quantification in these input parameters is crucial for validating nuclear reactor systems within the Best Estimate Plus Uncertainty (BEPU) framework. Inverse Uncertainty Quantification (IUQ) method has been used to quantify the uncertainty of PMPs from a Bayesian perspective. This paper introduces a novel hierarchical Bayesian model for IUQ which aims to mitigate two existing challenges: the high variability of PMPs under varying experimental conditions, and unknown model discrepancies or outliers causing over-fitting issues for the PMPs. 

The proposed hierarchical model is compared with the conventional single-level Bayesian model based on the PMPs in TRACE using the measured void fraction data in the Boiling Water Reactor Full-size Fine-mesh Bundle Test (BFBT) benchmark. A Hamiltonian Monte Carlo Method - No U-Turn Sampler (NUTS) is used for posterior sampling in the hierarchical structure. The results demonstrate the effectiveness of the proposed hierarchical structure in providing better estimates of the posterior distributions of PMPs and being less prone to over-fitting. The proposed hierarchical model also demonstrates a promising approach for generalizing IUQ to larger databases with a broad range of experimental conditions and different geometric setups.

\end{abstract}

\begin{keyword}
Hierarchical Bayesian Model \sep Inverse Uncertainty Quantification \sep Thermal Hydraulics \sep Markov Chain Monte Carlo \sep Surrogate Models 


\end{keyword}
\end{frontmatter}


\section{Introduction}
\label{introduction}

Best-estimate codes have been widely used in the nuclear system modeling and licensing process. These codes are designed to model all relevant physical processes in a realistic way and provide better insights into the accident progress. While a single calculation with a best-estimate code is considered the most accurate representation of reality, uncertainties in its results make it unsuitable for direct safety analysis. Consequently, Best-Estimate Plus Uncertainty (BEPU) methods are under fast development for the licensing process. BEPU approaches have been the focus of several OECD/NEA projects, such as UMS (\cite{amri2013overview}), BEMUSE (\cite{perez2011uncertainty}), etc. The results of these projects have shown that the uncertainty methods have a good maturity for the evaluation of the large-break loss-of-coolant accident (LB-LOCA) transients. 

As highlighted in the final BEMUSE project report (\cite{perez2011uncertainty}), significant efforts must be directed towards quantifying input uncertainty—an essential component of probabilistic uncertainty analysis in BEPU approaches.  Since most current BEPU methodologies rely on propagating uncertainties from input parameters to predictive model outputs, determining and justifying the uncertainty range associated with each uncertain parameter is crucial. Historically, the input uncertainty range is often determined by subjective expert judgment and therefore requires further development to provide a more scientific and rigorous method. This challenge was investigated in the international PREMIUM project (\cite{skorek2013premium}~\cite{mendizabal2016oecd}), while no consensus was reached by participants on guidelines and methodologies for input uncertainty quantification (UQ). The lack of consensus also motivated another OECD/NEA project SAPIUM (Systematic APproach for Input Uncertainty quantification Methodology) (\cite{baccou2020sapium}), which aims to develop a systematic approach for input UQ methodology in nuclear thermal-hydraulics (TH) codes.

The input uncertainties related to the numerical simulation of nuclear TH phenomena have multiple origins and can be classified into the different categories according to their origins (\cite{smith2013uncertainty}~\cite{barth2011brief}~\cite{wang2020hierarchical}):
\begin{itemize}
    \item Numerical uncertainties arising from the schemes employed in solving partial differential equations (PDEs). These uncertainties may originate from temporal and spatial discretization, errors due to iterative convergence of approximation algorithms, or round-off errors. Discretization uncertainty has been studied extensively and a number of techniques have been proposed for modeling this uncertainty.
    \item Uncertainties in the physical properties of materials and working fluids. Variations in properties such as density, specific heat, thermal conductivity, speed of sound, thermal expansion, and others can impact simulation outcomes. Accurate knowledge of these properties is crucial for the mathematical modeling and computer simulation of associated heat and mass transport processes.
    \item Uncertainties in geometries, initial/boundary conditions. The geometry information of relevant objectives in a model may not be exactly known to us. This type of uncertainty can be obtained from uncertainty in measurements or the manufacture tolerance. Initial and boundary conditions imposed to models can be caused by the inaccuracy in measuring them during experiment. The initial and boundary conditions may include the power level, pressure, temperature, flow rate, wall roughness, etc. 
    \item Uncertainties in physical phenomena. Uncertainty may arise due to lack of knowledge (epistemic uncertainty) in the formulation of the model or intentional simplifications of the model. In addition, empirical equations of state and constitutive equations (closure laws) utilized in two-fluid models or other fluid dynamics models may exhibit inherent probabilistic variations (aleatory uncertainty) in their parameters.
\end{itemize}

In this study, we focus on the uncertainties in physical phenomena present in the TRACE code (\cite{bajorek2008trace}),  which employs empirical correlations within the two-phase flow model to characterize transfer terms in balance equations. These correlations are obtained from different sets of experiments. However, their accuracy is subject to certain experimental conditions and may decrease due to different conditions. Thus, it is important to study the accuracy as well as the influence of those physical model correlations (\cite{wang2019inverse}). As has been shown in many benchmark studies on BEPU applications (\cite{glaeser2011bemuse}~\cite{wang2017sensitivity}), these uncertain parameters may have significant influences on Quantity-of-Interests (QoIs), thus it is important to improve our knowledge their uncertainties. Inverse UQ (sometimes also called input UQ) methods can be used to estimate the statistics of those parameters given experimental data, and the obtained statistics will useful for more accurate uncertainty and validation analysis.

Addressing input uncertainty distributions in this work can be broadly characterized as solving an inverse problem in modeling and simulation. This involves estimating model input parameters (with uncertainty) by comparing model outputs to experimental data. The inverse problem is an important approach to provide insights into parameters that cannot be directly observed. It has been extensively applied in various disciplines, including data assimilation, engineering optimization, medical imaging, and geophysics (\cite{tarantola2005inverse}). A common approach for addressing inverse problems in complex engineering systems is Bayesian inference. A comprehensive framework for input UQ for computer models, based on Bayesian formulation, was first introduced by \cite{kennedy2001bayesian}. In the filed of nuclear system TH, input UQ has mainly focused on the parametric uncertainty from closure equations in TH codes. \c{wu2021comprehensive} performed a comprehensive literature review on the IUQ methods for physical model parameters in nuclear system TH codes. This review paper summarized 12 existing IUQ methods from different research groups and compared them using various evaluation criteria. 

The landscape of IUQ has been considerably shaped by the progressive advancements in machine learning (ML) and artificial intelligence (AI). These technologies have significantly improved the precision and dependability of simulation models in various sectors, tackling a multitude of intricate challenges ~\cite{wang2023scientific}. Their successful deployment spans several disciplines, including healthcare (\cite{chen2019claims}), agriculture (\cite{wu2022optimizing}), reliability engineering (\cite{chen2020optimal, wang2024optimal, chen2017multi, chen2020some}), industrial engineering (\cite{chen2018data}), and artificial intelligence (\cite{liu2019dapred, liu2023stationary, liu2024cliqueparcel, wu2024switchtab, chen2023recontab}). The adaptability and success of ML/AI approaches in these areas provide compelling evidence of their potential and impart crucial learnings for advancing IUQ initiatives within the nuclear power industry.

Previous IUQ research in the nuclear engineering domain has primarily utilized \textit{single-level Bayesian inference} which uses Markov Chain Monte Carlo (MCMC) to explore the parameter posterior distributions. These approaches are often constrained to relatively small datasets. The calculated posterior distributions are valid only for the selected experimental cases and may vary when different datasets are used. To address this issue, this paper presents a \textit{hierarchical Bayesian modeling} approach. Hierarchical Bayesian models are particularly relevant when dealing with observations that are organized into distinct groups (\cite{gelman2013bayesian}). In such scenarios, calibration parameters may differ across groups, and employing traditional Bayesian inference methods may introduce substantial errors. Hierarchical models enable the definition and identification of ``hyperparameters'' to ensure that both ``group'' characteristics and ``individual'' characteristics are taken into account (\cite{wang2023scalable}).

A practical challenge for inverse problems is the large computational cost due to the long-running best-estimate codes, and the large number of samples needed in MCMC algorithms. Accurate and fast-running surrogate models can be used to significantly reduce the computational cost. The general idea is to train the surrogate models using a limited number of full model (the expensive computer code) simulations, usually with a certain learning algorithm. Many types of surrogate models have been developed, such as Polynomial Chaos Expansion (PCE) and Stochastic Collocation (SC) used by \cite{wu2017inverse}, Gaussian Process (GP) model by \cite{wu2018kriging} \cite{wang2019gaussian} and \cite{mc17-1}, Neural Networks by \cite{liu2022sam} and \cite{liu2021uncertainty}, etc. In this work, Polynomial Regression (PR) and GP are used as surrogate models and compared. The PR model might not be as accurate as other more complex surrogate models but it has proven to be sufficient for this work. The primary motivation for using PR is that it provides easily obtained gradient information of predictions, which is useful for the following gradient-based MCMC algorithm called No-U-Turn Sampler (NUTS) (\cite{hoffman2014no}). NUTS takes advantage of gradient information from the likelihood to achieve much faster convergence than traditional sampling methods, and it works well on high dimensional and complex posterior distributions (\cite{salvatier2016probabilistic}). The high-dimensional sampling space introduced by the hierarchical model requires the NUTS sampler to be used in this work.

The rest of the paper is organized as follows. Section \ref{sec2} will give an overview of IUQ method and its essential components. Section \ref{sec3} will introduce the hierarchical Bayesian framework and its details, and then in Section \ref{sec4}, the hierarchical framework is applied to a case study for TRACE physical model parameters using the BFBT benchmark steady-state void fraction data. Section \ref{sec5} will be the summary and conclusions.

\section{Overview of Inverse Uncertainty Quantification Methodology}
\label{sec2}

\subsection{Overview}

\begin{figure}[!h]
    \centering
    \includegraphics[width = \textwidth]{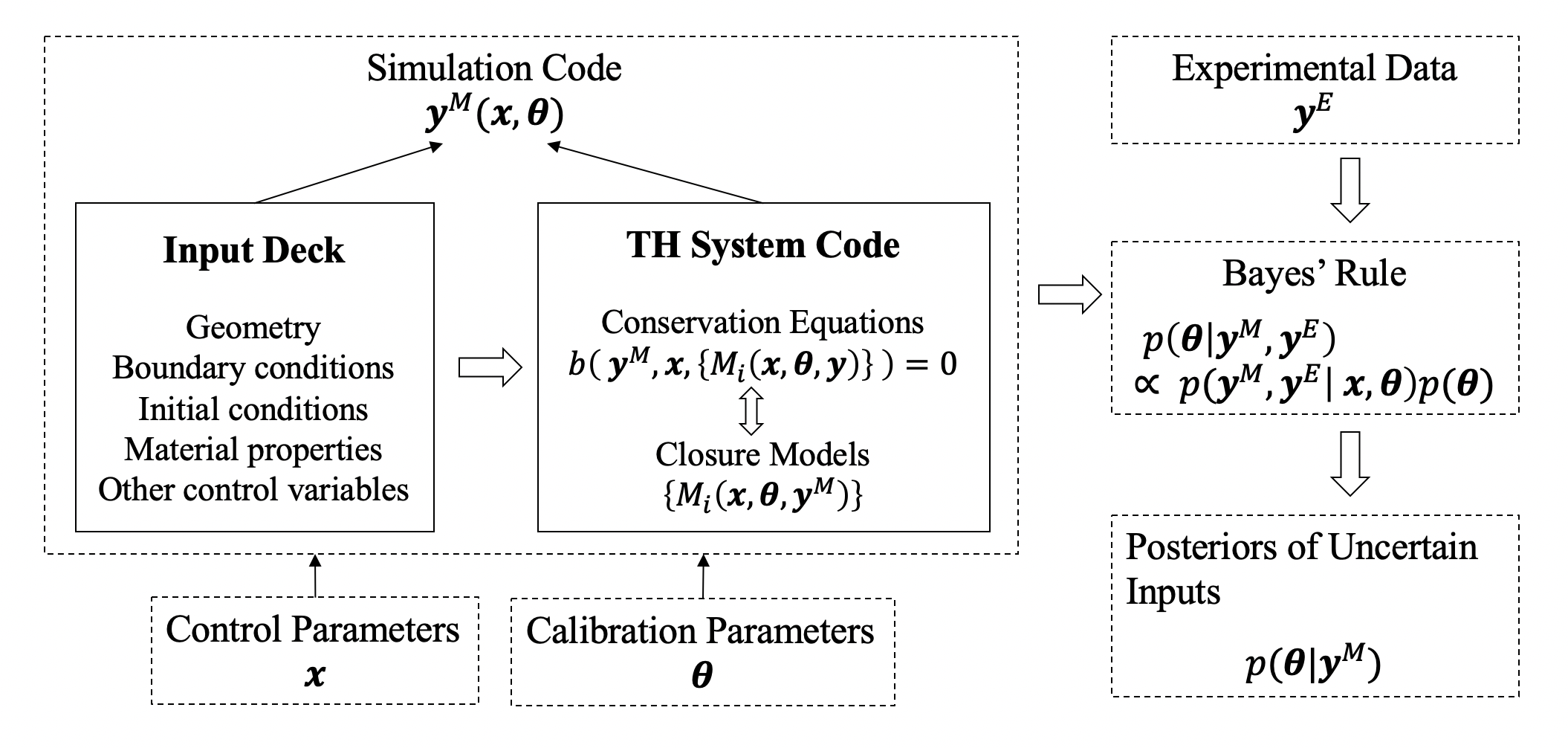}
    \caption{Major components in the IUQ framework}
    \label{fig:1-1}
\end{figure}

Figure \ref{fig:1-1} shows the major components used in the IUQ framework, where $\textbf{x}$ represents the control parameters, such as boundary conditions, initial conditions, etc. $\bm{\theta}$ represents calibration parameters, in this work it specifically represents the selected physical model parameters in the closure models of TH codes.

The input deck serves as the input for thermal-hydraulics (TH) codes, defining a specific system. It includes the geometrical configuration (i.e., nodalization), the materials and fluids involved, the initial and boundary conditions, and potentially the settings for the numerical solver. Control parameters $\textbf{x}$ are selected from those specifications. The TH simulation codes TRACE consists of 6 conservation equations, which are then closed with additional set of closure models $M_i(\bm{x,\theta,y^M})$ (\cite{wicaksono2018bayesian}). The output of the simulation code $\bm y^M$ can be combined with corresponding experimental data $\bm y^E$ in the Bayes' rule, resulting in a joint posterior probability density distribution (PDF) for the selected inputs.

The detailed procedures of the IUQ framework can be organized into the following stages: 
\begin{enumerate}
    \item Define the problem under investigation and select appropriate experimental data and simulation codes. In this study, the BFBT (Boiling Water Reactor Full-size Fine-mesh Bundle Test) benchmark is used as the basis, with corresponding TRACE models developed.
    \item Determine influential input parameters within the chosen TH codes.
    \item Conduct a two-step sensitivity analysis to identify significant input parameters. Initially, implement a relatively straightforward perturbation method to all parameters, selecting those that are active in the model. Subsequently, employ a more precise sensitivity analysis technique, Sobol indices, to identify the influential parameters. Surrogate models may be used in this stage.
    \item Develop a hierarchical Bayesian inference model to quantify the posteriors of uncertain inputs. This stage integrates experimental data with simulation outcomes to establish input distributions that result in improvement agreement with experimental data. Surrogate model and MCMC sampling algorithms are used in this stage.
    \item Validate the derived posteriors.
\end{enumerate}

\subsection{Model Updating Equation}

A key assumption in the Bayesian IUQ framework is the model updating equation. Following the work of \cite{kennedy2001bayesian}, we represent the relationship between the computer model outputs $\bm y^M(\bm x, \bm \theta)$ and the observations $\bm y^E(\bm x)$ in the equation:

\begin{equation}
\label{eqa:21}
    \bm y^E(\bm x) = \bm y^M(\bm x, \bm \theta) + \bm \delta(\bm x) + \bm \epsilon 
\end{equation}
where $\bm \epsilon $ is the observation error that is usually assumed to be independent and identically distributed as $\mathcal{N}(0,\sigma_{exp}^2)$. It should be noted that this assumption may not always hold in reality, especially in time-dependent problems (\cite{wangsurrogate}). $\bm \delta(\bm x)$ is the model discrepancy term, which is caused by incomplete or inaccurate physics employed in the model. Here, the parametric uncertainty is derived from the $\bm \theta$ parameter, and other forms of uncertainties are incorporated in the model discrepancy term. 

Following the model updating equation, the posterior PDF of the calibration parameters can be found using Bayes' rule:

\begin{equation}
\label{eqa:22}
    p(\bm \theta | \bm y^E, \bm y^M) \propto p(\bm y^E, \bm y^M |\bm \theta) \cdot p(\bm \theta)
\end{equation}
where $p(\bm \theta)$ is the prior distribution of the calibration parameters, and $p(\bm y^E, \bm y^M |\bm \theta)$ is the likelihood function. From equation \ref{eqa:21}, we know that $\bm \epsilon = \bm y^E(\bm x) - \bm y^M(\bm x, \bm \theta) - \bm \delta(\bm x) $ follows a multivariate normal distribution. So the posterior can be written as:

\begin{equation}
    \label{eqa:23}
    p(\bm \theta | \bm y^E, \bm y^M) \propto \frac{1}{\sqrt{|\bm \Sigma_t|}} \exp \bigg[-\frac{1}{2}[\bm y^E-\bm y^M - \bm \delta]^T \bm{\Sigma}_t^{-1}[\bm y^E-\bm y^M - \bm \delta] \bigg] \cdot p(\bm \theta)
\end{equation}

The covariance matrix $\bm{\Sigma}_t$ is defined as: 
\begin{equation}
    \label{eqa:sigma_total}
    \bm{\Sigma}_t = \bm \Sigma_{exp} + \bm \Sigma_{\delta} + \bm \Sigma_{code}
\end{equation}
where $\bm \Sigma_{exp}$ is the experimental uncertainty, $\bm \Sigma_{\delta}$ is the model uncertainty, and $\bm \Sigma_{code}$ is the code uncertainty, introduced by using surrogate models to replace the full model. $p(\bm \theta)$ is the prior distribution of calibration parameters and can be treated as a non-informative uniform distribution over a certain range.  

It can be very challenging to estimate the model discrepancy $\bm \delta$. In cases where model discrepancy is negligible, we can simply ignore the $\bm \delta$ term. However, ignoring the model discrepancy when it does exist can cause over-fitting issue of the calibration parameters, meaning that the IUQ process will favor a $\bm \theta $ distribution that yields the best agreement between the measurement data and the model simulation, rather than the ``true'' value. A method called the Modular Bayesian Approach (MBA) was developed to tackle the model discrepancy problem in IUQ (\cite{wu2018inverse}~\cite{wu2018inverse22}~\cite{wang2018ans}).

\subsection{Sensitivity Analysis}

One question we need to answer at the very beginning of IUQ is the identification of input parameters requiring study. Sensitivity Analysis (SA) examines how uncertain inputs contribute to variations in model QoIs. There are typically two classes of SA: local SA and global SA (\cite{saltelli2002making}). Local SA addresses the local impact of input variations, whereas global SA considers the entire variation range of inputs and can also provide insights into how interactions between input parameters influence QoIs. Generally, local SA is computationally less expensive than global SA; thus, in the IUQ framework, we use local SA for initial parameter selection and global SA for a more accurate study to determine the final list of inputs to be investigated.

Sobol indices method provides a straightforward measure of sensitivity in arbitrarily complex computational models. It is a variance-based method where the variance of the output is decomposed as a sum of contributions of each input variable and their combinations. This decomposition of variance is referred to as ANOVA (ANalysis Of VAriance). Sobol indices method has been successfully applied to many UQ related applications (\cite{aly2019variance}~\cite{wang2017sensitivity}~\cite{mc17-2}) and the details of the method will not be repeated in this paper. The calculation of Sobol indices in this paper will follow the samping-based approach outlined by \cite{saltelli2010variance}. This approach requires $N(d+2)$ number of samples. $N$ is usually a value of several thousand for sufficient accuracy and $d$ is the dimension of the input, and GP surrogate model is used here due to a large number of code runs required.

\subsection{Surrogate Models}

Surrogate models, also known as metamodels or emulators, play a critical role in Bayesian calibration of computer models, particularly when dealing with computationally demanding simulations such as nuclear reactor analyses. These surrogate models are constructed using a limited number of full model runs at specifically chosen input parameter values, combined with a learning algorithm. Many types of surrogate models have been utilized in the Bayesian calibration process,  Different types of surrogate models have different characteristics, so they may fit in different engineering scenarios. Factors to consider when selecting an appropriate surrogate model include input dimensionality, the non-linearity of the input-output relationship, and whether the problem is time-dependent or steady-state. 

In this work, we primarily employ two types of surrogate models: GP and PR. GP has been widely used as surrogate models since the seminal work of Bayesian calibration by \cite{kennedy2001bayesian} utilizing GP modeling. One notably characteristic of GP is that it not only provides a mean estimator at any untried input location, but also the corresponding variance estimator, which can serve as the code uncertainty term ($\bm \Sigma_{code}$) in Equation \ref{eqa:sigma_total}. In other surrogate models, since this term is not available, we will need to treat the covariance term as a calibration parameter. Details of the GP surrogate model will not be repeated in this paper, interested readers may refer to \cite{rasmussen2004gaussian} for more details on GP and \cite{wu2018inverse}~\cite{wang2019gaussian} for similar IUQ applications using GP.

In addition to the GP model, as a more convenient but less accurate alternative, Polynomial Regression model is also used in the current work. The incentive of using PR derives from the requirement of gradient information in some advanced and efficient MCMC algorithms. As a parametric regression model, it is straightforward to calculate the gradient at any given location. The PR model with degree of $d$ for two variables $x_1,x_2$ has the following form:
\begin{equation}
    \hat{y}(w,x) = w_0 + w_1x_1 + w_2x_2 + w_3x_1x_2 + w_4x_1^2 + w_5x_2^2 + ... +w_nx_1^d + w_{n+1}x_2^d
\end{equation}
where all d\textsuperscript{th} order polynomials and interaction terms are considered. The parameter $\bm w = [w_1, w_2, ...]$ can be easily estimated by Ordinary Least Squares method, similar to other linear regression models. 

The accuracy of the surrogate models needs to be assessed before use. We are particularly interested in the predictive accuracy at untried points, which can be achieved by quantifying the predictive error using an additional set of validation data. Various metrics, such as the Mean Squared Error, Leave-One-Out-Error, Coefficient of Determination, and the Mean Absolute Error, can be employed to measure the discrepancy between the predicted values and the actual values. By comparing these metrics, we can ascertain the effectiveness of the surrogate model in capturing the underlying patterns and relationships in the data, ensuring its reliability for subsequent analysis and applications.

\section{Hierarchical Bayesian Framework for Inverse Uncertainty Quantification}
\label{sec3}

Prior IUQ research in the field of nuclear engineering primarily employs single-level Bayesian inference models and MCMC algorithms to calculate the posterior distributions of specific parameters. However, these methods are often limited to relatively small datasets. The derived posterior distributions are only applicable to the chosen data and may fluctuate given different datasets due to the fact that the calibrated parameters are over-fitted to the selected data. This paper presents a hierarchical Bayesian model to address this limitation. The hierarchical Bayesian model is particularly relevant when we have observations that are organized in groups in which the calibration parameters may differ.

\subsection{Hierarchical Bayesian Model}

\begin{figure}[!htbp]
    \centering
    \includegraphics[width = \textwidth]{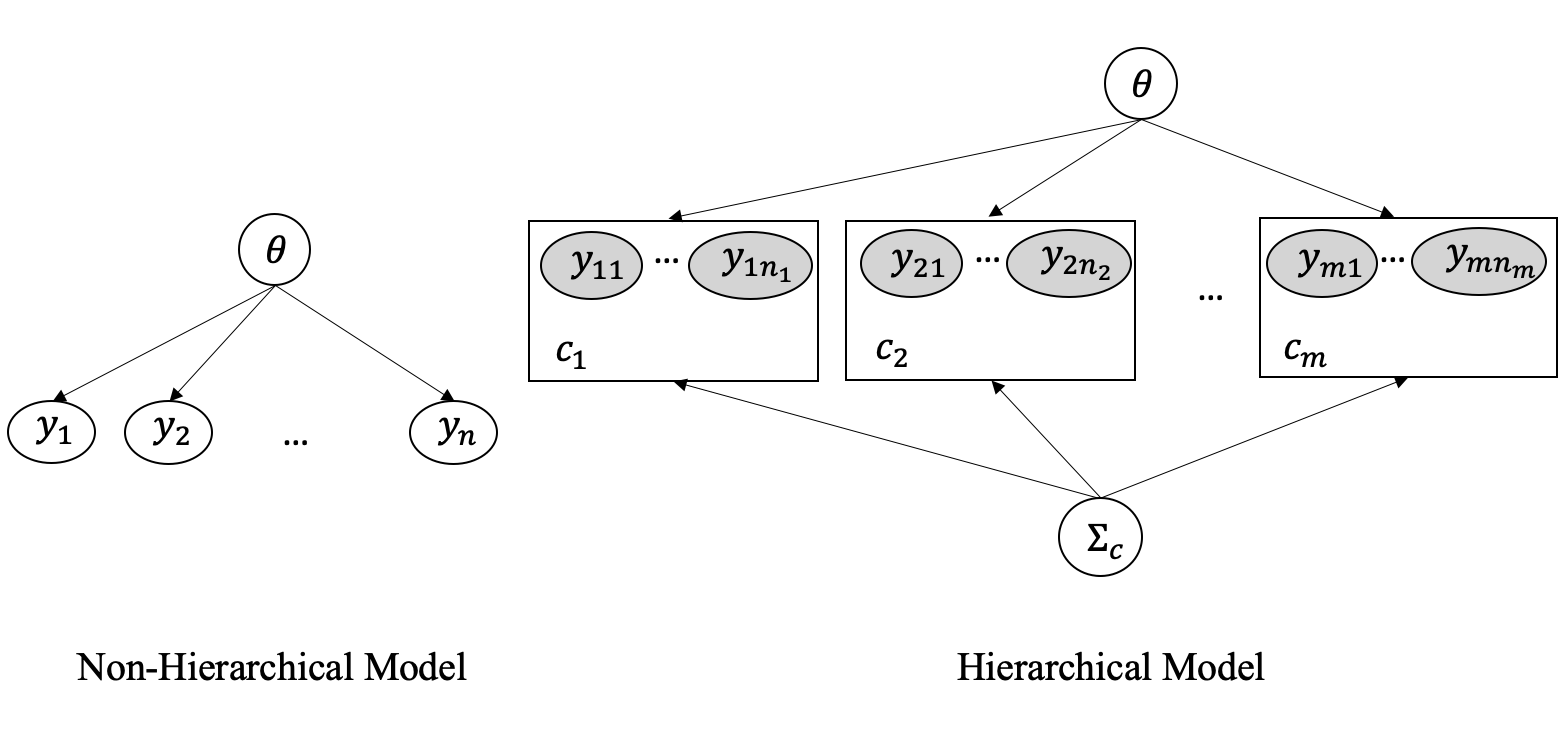}
    \caption{Illustration of the non-hierarchical and hierarchical models.}
    \label{fig:hb1}
\end{figure}

In the traditional Bayesian calibration setting, we have assumed that the observations are independent of each other so that the joint likelihood function can easily be formulated as the product of each individual likelihood. However, in many situations, such independence may not hold. Hierarchical model is also called a multi-level model or mixed-effect model. A simple illustration of the structure of the hierarchical model is shown in Figure \ref{fig:hb1}. Observations $\bm y$ are from different clusters determined by (1) cluster parameter $c_m$ which might be different among clusters, and (2) parameter $\bm \theta$ shared across clusters. If the probability distribution of $c_m$ can be parameterized by $\bm \Sigma_c$, $\bm \Sigma_c$ can be seen as the cluster-specific parameter, or per-group parameter, to be estimated.

Consider the simple example illustrated in Figure \ref{fig:hb1}, where the observations are assumed to follow a normal distribution. In this scenario, the parameters to be estimated are the mean and variance. Let's assume that all clusters share the same variance, denoted as $\sigma^2$, but have different means $\mu_i$ for each cluster. So the $\sigma^2$ is a shared parameter ($\theta$ in Figure \ref{fig:hb1}) and $\mu_i$ is per-group parameter ($c_i$ in Figure \ref{fig:hb1}). Now we would like to specify the distribution over the cluster-specific $\mu_i$. We can assume the distributions are also normal, or other task-specification distributions, without loss of generality. If normal distributions are assumed, two parameters (global mean $\mu$ and global standard deviation $\sigma_y$, correspond to the $\Sigma_c$ in Figure \ref{fig:hb1}) would be required. So we can describe the parameters $\mu_i$ and observations $y_{ij}$ as:
$$
\mu_i \sim N(\mu, \sigma_y^2)
$$
$$
y_{ij} \sim N(\mu_i, \sigma^2 )
$$

\begin{figure}[!htbp]
    \centering
    \includegraphics[width = 0.8\textwidth]{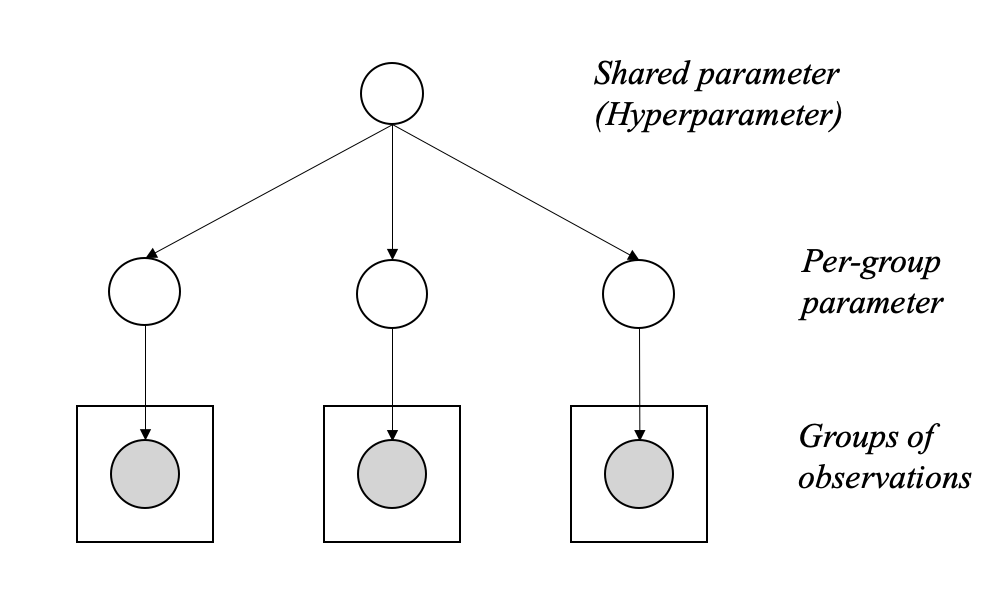}
    \caption{Illustration of a classical hierarchical model structure.}
    \label{fig:hb2}
\end{figure}

Now we can re-arrange the structure of the above hierarchical model to better illustrate the process of the two preceding equations: the per-group parameters are generated from shared parameters, and the observations are generated from per-group parameters. This relationship is shown in Figure \ref{fig:hb2}. 

In summary, the application of hierarchical models is highly beneficial when there's an observable existence of group-level characteristics, and the data's structure can be effectively modeled or assumed. When dealing with hierarchically structured data, utilizing a non-hierarchical model may not be appropriate for two reasons: (1) a model containing insufficient parameters may not successfully fit the dataset, and (2) a model with an abundance of parameters may cause over-fitting of the dataset. In contrast, hierarchical models, having a substantial number of parameters for fitting the data accurately and a population distribution to model the inter-dependencies of the parameters, can avoid the issue of over-fitting. This is achieved as the parameters are ``constrained'' by the population distribution, preventing them from exactly replicating the data.

\subsection{Bayesian Inference in Hierarchical Models}

\begin{figure}[!htbp]
    \centering
    \includegraphics[width = 0.4\textwidth]{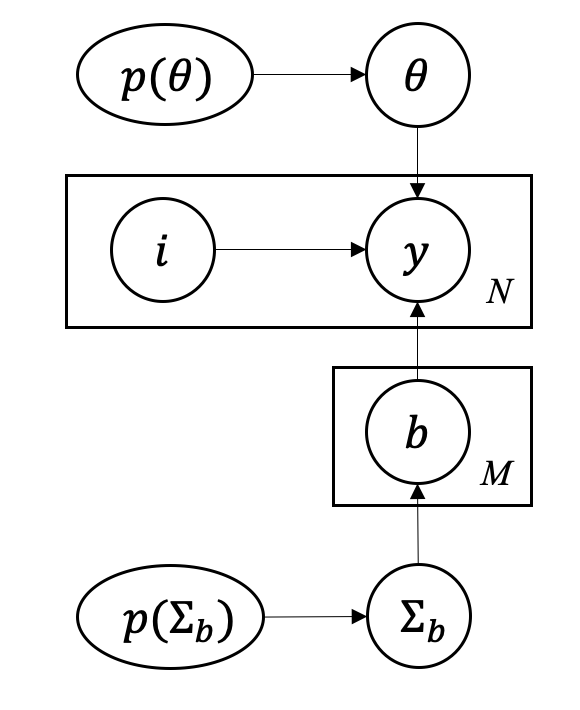}
    \caption{Graphical representation of a hierarchical model.}
    \label{fig:hb3}
\end{figure}

In this section we describe how to estimate parameters in a hierarchical Bayesian model structure. The hierarchical model structure in Figure \ref{fig:hb1} can be plotted into a more succinct and formal graphical representation, as shown in Figure \ref{fig:hb3}. In the figure, $p(\bm \theta)$ is the prior distribution of $\bm \theta$, $i$ is the cluster index, $\bm y$ is the observation, and there are $N$ observations in total. The per-group parameter $b_i$ is governed by its distribution $\bm {\Sigma_b}$, which has a prior distribution $p(\bm {\Sigma_b})$.\\

It should be noted that within this model, our primary goal is to learn about $\bm \theta$ and $\bm {\Sigma_b}$, as opposed to the individual group-specific parameters $b_i$, but $b_i$ also has to be estimated in order to estimate $\bm{\Sigma_b}$. So we need to marginalize over the $b_i$ parameters after we have obtained the joint posterior distributions. There are generally two approaches to estimate the posteriors, we can either use Maximum Likelihood Estimation (MLE) to obtain best-fit point estimations or use sampling to get full posterior distributions for these parameters. 

In the MLE method, the likelihood of $\theta$ and $\Sigma_b$, marginalized over $b$, can be expressed as:
\begin{equation}
    L(\bm {\Sigma_b}, \bm \theta; \bm y) = \int_b P(\bm y \mid \bm \theta, \bm b, i)P(\bm b|\bm {\Sigma_b}) \,db
\end{equation}

In case full posterior distributions are desired, we can integrate the prior distribution and apply Bayes' rule. The prior distribution can be decomposed into the following form:

\begin{equation}
    P(\bm {\Sigma_b}, \bm \theta, b_i) = P(b_i \mid \bm {\Sigma_b}) \cdot P(\bm {\Sigma_b}, \bm \theta)
\end{equation}
which is based on the fact that $\bm {\Sigma_b}$ influences observations only through $b_i$. So the marginalized joint posterior distribution can be expressed as:
\begin{equation}
    P(\bm {\Sigma_b}, \bm \theta \mid \bm y) \propto \int_b P(\bm y \mid \bm \theta, \bm b, i)P(\bm b \mid \bm {\Sigma_b}) \cdot P(\bm \theta)P(\bm {\Sigma_b}) \,d b
\end{equation}

Hierarchical models provide a flexible framework for modeling the complex interactions but come with higher computational requirements in inference. The parameter space to be estimated can be high-dimensional: we need to estimate cluster-specific parameter $b_i$, which may contain up to hundreds. Traditional random-walk based MCMC algorithms such as Metropolis-Hastings (MH) or advanced MH algorithms tailored for relatively high dimensional problems still suffer from serious convergence issues because that the random behavior of the proposal function in very inefficient in high-dimension domains. Hamiltonian Monte Carlo (HMC) and No-U-Turn Sampler (NUTS), which will be introduced in the following section, would be useful in such high-dimensional conditions.

\subsection{Markov Chain Monte Carlo sampling}

Estimating the posterior in Equation \ref{eqa:23} requires numerical sampling method as analytical solution is not available. Statistical tools that can be used here include MLE, Maximum A Posteriori (MAP), Variational Inference (VI), Laplace Approximation (LA), and various MCMC methods. MLE, MAP, LA all give point estimates and VI, LA are both approximate solutions rather than exact solutions. MCMC, although more time consuming, can estimate the exact posteriors for arbitrary distributions. It is a class of algorithms for sampling from an arbitrary probability distribution, for example, the probability distribution of equation \ref{eqa:23}, where the density function needs to be known only up to a normalizing constant. 

The MH algorithm is one of the most widely used MCMC methods for sampling from a target distribution. In this approach, a family of potential transitions from one state to the next is determined using a proposal distribution. However, when the number of parameters to be estimated increases to 20 or 30, the random walk-based methods, such as the traditional MH algorithm, suffer from a very low acceptance rate. To address these limitations, the HMC method has been developed as an alternative to the conventional MH algorithm. HMC is an MCMC technique that avoids the random walk behavior and the sensitivity to correlated parameters by utilizing first-order gradient information to inform the sequence of steps taken during the sampling process (\cite{andrieu2008tutorial}~\cite{hoffman2014no}). By employing an auxiliary variable scheme, HMC can transform the problem of sampling from a target distribution into simulating Hamiltonian dynamics, effectively suppressing the random walk behavior. As a result, HMC offers a more efficient approach to sampling from high-dimensional distributions and provides a smoother transition between states, thus enhancing the performance of MCMC methods in scenarios with a large number of parameters to be estimated.

In HMC, an auxiliary momentum variable $r_d$ is introduced for each model model parameter $\theta_d$. In most cases, these momentum variables are drawn independently from standard normal distribution, yielding the joint density $p(\theta,r) \propto \exp(L(\theta) - \frac{1}{2}r \cdot r)$, where $L(\cdot)$ is the logarithm of the joint density of the variables of interest $\theta$. We can interpret this  model in physical terms as a fictitious Hamiltonian system where $\theta$ is a particle’s position in D-dimensional space, $r_d$ is the momentum of that particle in the d-th dimension, $L$ is a position-dependent negative potential energy function, $\frac{1}{2}r \cdot r$ is the kinetic energy of the particle, and $log(p(\theta,r))$ is the negative energy of the particle. The evolution of the Hamiltonian dynamics of this system can be simulated via the ``leapfrog” integrator (\cite{hoffman2014no}), which will update the $r$ and $\theta$ accordingly. The details of the algorithm is shown in Algorithm \ref{algo:2}.

\begin{algorithm}[ht]
\SetAlgoLined
\caption{Hamiltonian Monte Carlo Algorithm (\cite{andrieu2008tutorial}~\cite{kass1998markov})}
\label{algo:2}
\textbf{For} m = 1, 2, ..., M:\\
\qquad Sample $r^0 \sim N(0, I)$.\\
\qquad Set $\theta^m  = \theta^{m-1}, \Tilde{r} = r^0 $.\\
\qquad \textbf{For} i = 1, 2, ... , $l$:\\ 
\qquad \qquad Set $\Tilde{\theta},\Tilde{r}$ = Leapfrog($\Tilde{\theta}, \Tilde{r}, \epsilon$).\\
\qquad \textbf{End} for\\
\qquad Compute
$$\alpha = \min(1, \frac{\exp(L(\Tilde{\theta}) - 0.5 \Tilde{r} \cdot \Tilde{r})}{\exp(L(\theta^{m-1}) - 0.5 r^0 \cdot r^0)})   $$
\qquad With probability $\alpha$, set $\theta^m = \Tilde{\theta}$,$r^m = \Tilde{r}$\\ 
\textbf{End} for\\
\vspace{4mm}
\textbf{Function} Leapfrog($\theta, r, \epsilon$):\\
$$\Tilde{r}  = r + \frac{\epsilon}{2}\nabla_\theta L(\theta)$$
$$\Tilde{\theta}  = \theta + \epsilon\Tilde{r}$$
$$\Tilde{r}  = \Tilde{r} + \frac{\epsilon}{2}\nabla_\theta L(\Tilde{\theta})$$
\qquad \textbf{return} $\Tilde{\theta}, \Tilde{r}$.
\end{algorithm}

Suppose we are drawing $M$ samples using the HMC algorithm. For each sample, we first re-sample the momentum variables from $N(0,I)$, which can be interpreted as a Gibbs sampling update. Then, the leapfrog function is used $l$ times to update the variables $\theta$ and $r$, and a proposal position-momentum pair of $\Tilde{\theta}$ and $\Tilde{r}$ is generated. The proposal can be accepted or rejected according to the Metropolis algorithm. The leapfrog functions plays an important role in improving the efficiency of sampling. By taking $l$ leapfrog steps per sample, the proposal generated for $\theta$ has a high probability of being accepted. However, as we can see from the definition of leapfrog function, derivative of the joint density function of $\theta$ is required. The performance of the HMC also depends on hand-chosen values of $\epsilon$ and $l$. If the two parameters are not properly selected, the Markov chain may be slow to move between regions of high and low densities.

NUTS is an extension of HMC that eliminates the need to manually tune these parameters in HMC. The idea is to use a recursive algorithm to build a set of likely candidate points that spans the target distribution, and stop automatically when it starts to retrace its steps. A detailed description about NUTS can be found in \cite{hoffman2014no}. In this work, the NUTS is implemented with an open-source package PyMC3.8 (\cite{salvatier2016probabilistic}). A significant advantage of NUTS is that it can be used without any hand-tuning, making it suitable for a wide range of engineering applications. However, as can be seen in the algorithm, the gradient of the log-likelihood with respect to the sampling parameter is required. Therefore, the surrogate model should also be capable of providing the gradient of outputs if NUTS is to be employed.

\subsection{Application to Synthetic Data}
\subsubsection{Problem Definition}
The synthetic data example is constructed in a way that is similar to the thermal-hydraulics system code application using BFBT void fraction data, which we will be utilizing in a subsequent sections. Suppose there are three parameters to be estimated $[\alpha,\beta,\theta]$, and we know that observations are from a quadratic function ($X$ is determined):
\begin{equation*}
    y_{obs}=\alpha X^2 + \beta X + \theta + \mathcal{N}(0,\sigma^2) 
\end{equation*}
and there 100 groups of data, $i = 1,2,…,100$, each group contains 5 data, $j = 1,2,3,4,5$. The true values of $[\alpha,\beta,\theta]$ have slight variability among different groups, they are from a common normal distribution:
\begin{align*}
    \alpha_i &\sim \mathcal{N}(4,1) \\
    \beta_i &\sim \mathcal{N}(2,1) \\
    \theta_i &\sim \mathcal{N}(-2,1)
\end{align*}

Now the objective of this problem is to estimate the above distributions of $[\alpha,\beta,\theta]$, only knowing $X,y_{obs}$ and the quadratic relationship between $X$ and $y_{obs}$. Conceptually, $[\alpha,\beta,\theta]$ can be considered as the physical model parameters (PMPs), while the quadratic relationship can be likened to the complex simulations of nuclear system codes.

\subsubsection{Hierarchical Structure}
\begin{figure}[!htbp]
    \centering
    \includegraphics[width = 0.5\textwidth]{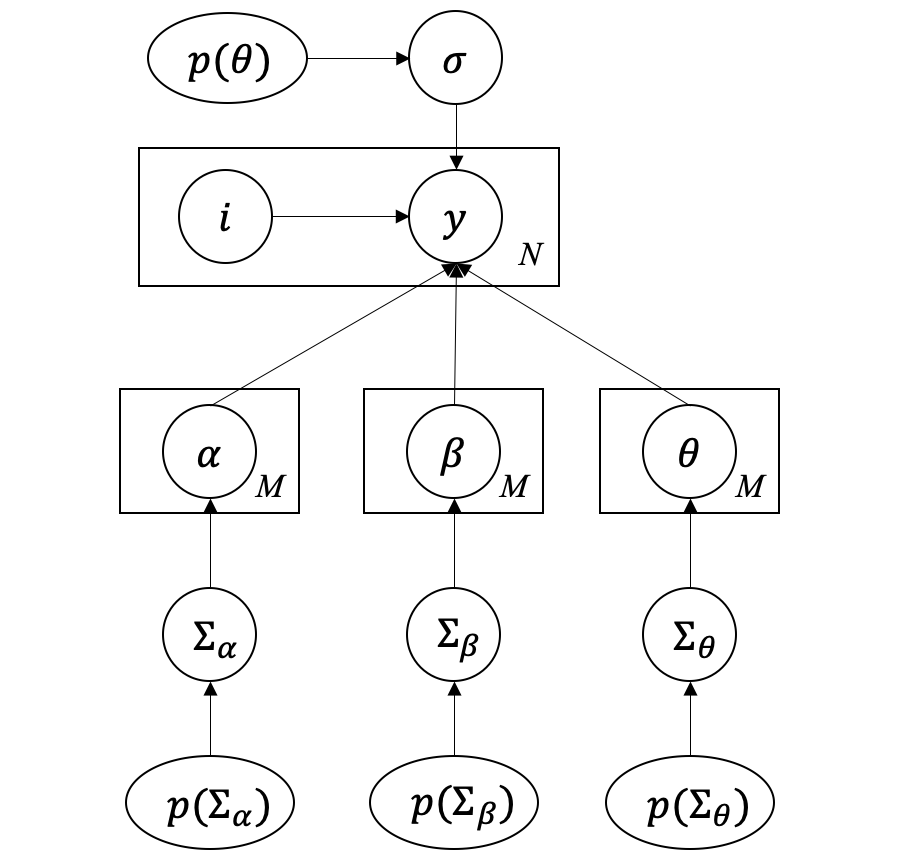}
    \caption{Graphical representation of the hierarchical model structure for the toy example}
    \label{fig:hb4}
\end{figure}

Given that the parameters $(\alpha,\beta,\theta)$ differ across groups, a Bayesian hierarchical model is employed to address this issue. The hyperparameters of $(\alpha,\beta,\theta)$ are defined as $(\mu_{\alpha},\mu_{\beta},\mu_{\theta},\sigma_{\alpha},\sigma_{\beta},\sigma_{\theta})$. A graphical representation of this model structure can be seen in Figure \ref{fig:hb4}. $\Sigma_{\alpha}$ denotes the hyperparameters of $\alpha$, which is $\mu_{\alpha}$ and $\sigma_{\alpha}$. Same notations apply for $\beta$ and $\theta$. Consequently, we have a set of shared parameters ($\sigma, \Sigma_{\alpha},\Sigma_{\beta},\Sigma_{\theta}$) and cluster-specific parameters ($\alpha_i, \beta_i, \theta_i$). Relatively wide uniform priors are imposed on those shared parameters:

\begin{align*}
    \mu_{\alpha} & \sim \mathbf{Unif}(-10,10) \\
    \sigma_{\alpha} & \sim \mathbf{Unif}(0,10) \\
    \mu_{\beta} & \sim \mathbf{Unif}(-10,10) \\
    \sigma_{\beta} & \sim \mathbf{Unif}(0,10) \\
    \mu_{\theta} &\sim \mathbf{Unif}(-10,10) \\
    \sigma_{\theta} &\sim \mathbf{Unif}(0,10)\\
    \sigma & \sim \mathbf{Unif}(0,10)
\end{align*}

The input parameters $\alpha_i,\beta_i,\theta_i$ can then be sampled from the normal distribution defined by $(\mu_{\alpha},\mu_{\beta},\mu_{\theta},\sigma_{\alpha},\sigma_{\beta},\sigma_{\theta})$:

\begin{align*}
    \alpha_i &\sim \mathcal{N}(\mu_{\alpha},\sigma_{\alpha}) \\
    \beta_i &\sim \mathcal{N}(\mu_{\beta},\sigma_{\beta}) \\
    \theta_i &\sim \mathcal{N}(\mu_{\theta}, \sigma_{\theta}) 
\end{align*}

In this formulation, we are faced with the challenge of estimating $307 (100*3 +6 + 1)$ parameters in the Bayesian inference process. Utilizing random sampling-based MCMC algorithms would be impractical due to the high dimensionality of this problem. Therefore, to address this challenge, the NUTS method is employed using the PyMC3.8 package (\cite{salvatier2016probabilistic}). 

\subsubsection{Results and Discussions}
\begin{figure}[!h]
    \centering
    \includegraphics[width = 0.8\textwidth]{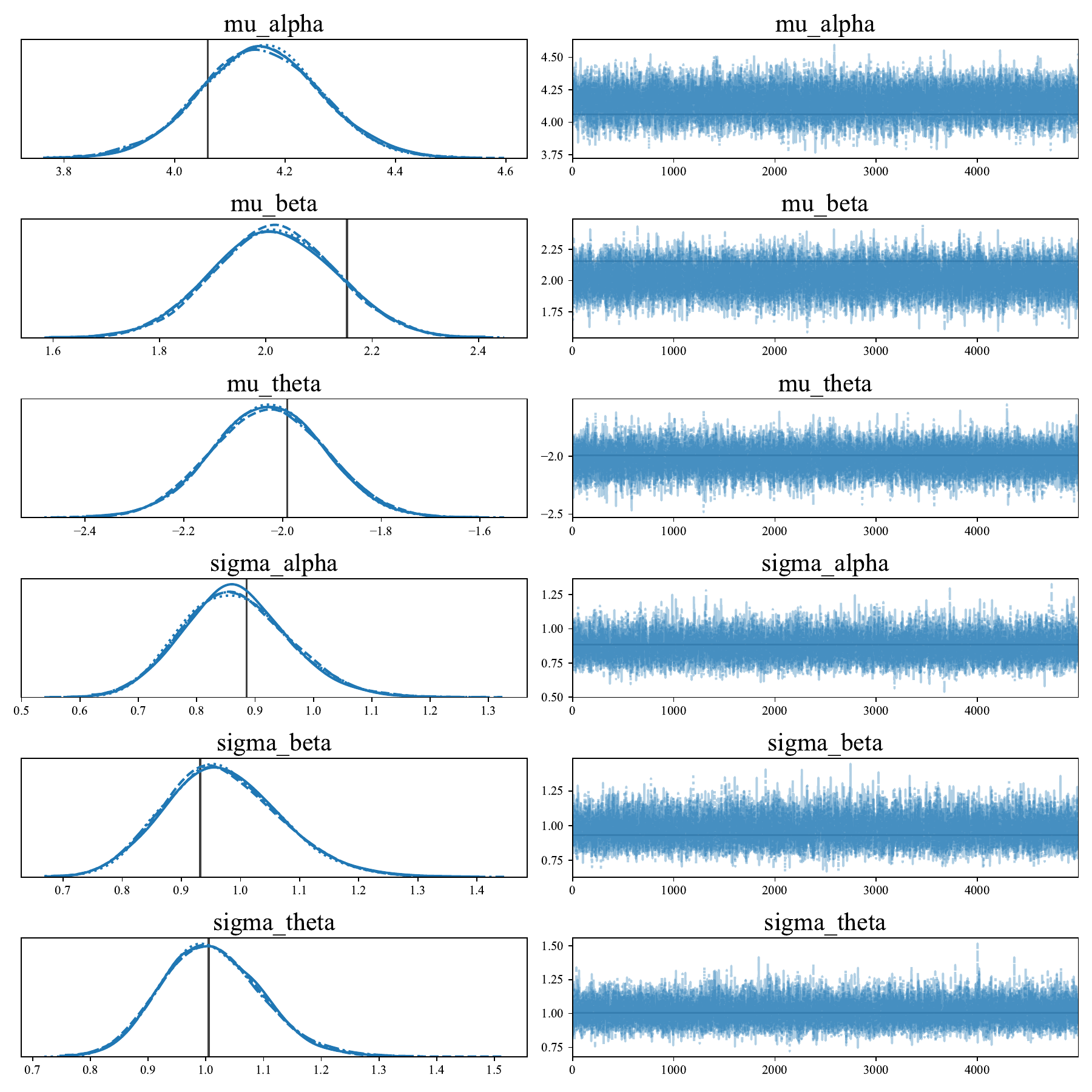}
    \caption{Posterior distributions of hyper-parameters and their trace plots}
    \label{fig:toy1}
\end{figure}

The posterior distributions of the hyper-parameters sampled by NUTS are shown in Figure \ref{fig:toy1}. The vertical lines on the left side are the true values from the samples, and the figures on the right are the sampling trace plots. These trace plots demonstrate well-mixed chains, and the consistency across the four parallel sampling chains is obvious, as represented by the different line styles in the left figures. Although a minor discrepancy can be observed for the first two parameters, the true parameters remain within the corresponding probability distributions.

\begin{figure}[!htbp]
    \centering
    \includegraphics[width = \textwidth]{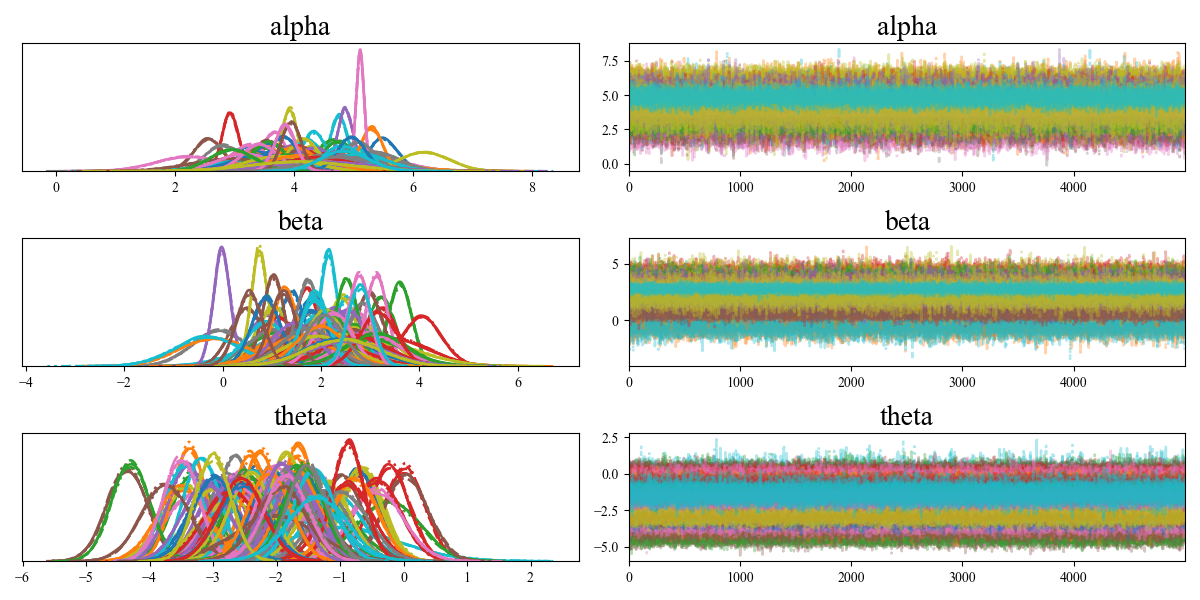}
    \caption{Posterior distributions of cluster-specific parameters and their trace plots.}
    \label{fig:toy3}
\end{figure}

It is also interesting to see the calibration results for all cluster-specific parameters. Figure \ref{fig:toy3} shows all the posterior distributions for each group's $\alpha_i, \beta_i, \theta_i$ as well as their trace plots. It can be seen that these individual parameters do have relatively large differences. It can be reasonably inferred that if we were to build a model with many group-specific parameters, the model would ``over-fit'' since the estimated parameters will only be valid in certain groups. If we were to use a simple non-hierarchical model, the model would not be able to capture true distributions of the calibration parameters.

Figure \ref{fig:toy3} provides a closer examination of the IUQ results for all cluster-specific parameters, showcasing all posterior distributions for each group's $\alpha_i, \beta_i, \theta_i$ along with their trace plots. It is clear that these individual parameters exhibit considerable variability. This variability suggests that, if we constructed a model with numerous group-specific parameters, such a model would ``over-fit``, as the estimated parameters would only be valid within certain groups. Consequently, a simple non-hierarchical model would not be able to capture the true distributions of the calibration parameters accurately.

\begin{figure}[!htbp]
    \centering
    \includegraphics[width = 0.7\textwidth]{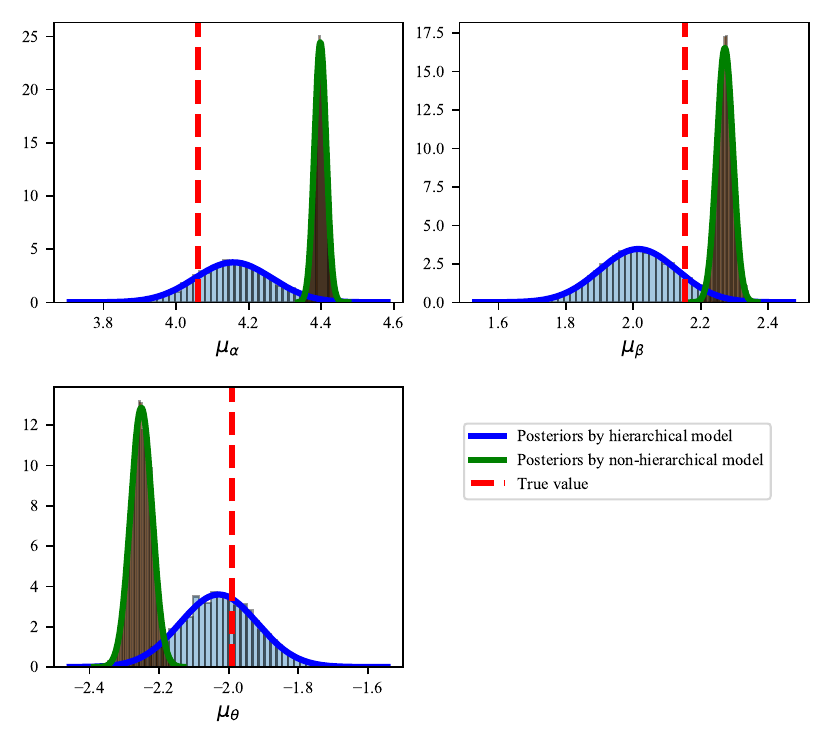}
    \caption{Posterior distributions by hierarchical model and non-hierarchical model.}
    \label{fig:toy2}
\end{figure}

For comparison purposes, a simple non-hierarchical model for this toy example is also constructed. In this non-hierarchical model, three global parameters $\alpha, \beta$, and $\theta$ are estimated and no groups are involved. A comparison between the posterior distributions obtained by the hierarchical model and the non-hierarchical model is shown in Figure \ref{fig:toy2}. 

We can see that the non-hierarchical model struggles to handle the variability of the calibration parameters, resulting in inaccurate estimations. In contrast, the hierarchical model successfully includes the true values in its posterior distributions. This comparison demonstrates the effectiveness of hierarchical models when dealing with data that exhibits grouping or clustering.
\section{Application to Thermal-Hydraulics System Code}
\label{sec4}

In this section, we will apply the hierarchical Bayesian framework to IUQ problem for the PMPs in TRACE code, using the BFBT benchmark dataset as a practical case study.

\subsection{BFBT Benchmark and TRACE simulations}

The international OECD/NRC BWR full-size fine-mesh bundle test (BFBT) benchmark (\cite{neykov2006nupec}) was created to encourage advancement in sub-channel analysis of two-phase flow in rod bundles, which has great relevance to the nuclear reactor safety evaluation. The BFBT test program includes experiments performed under steady-state and transient conditions, with measurements for single and two-phase pressure losses, void fraction, and critical power. The benchmark has been widely used in the validation process of various computational approaches (\cite{wang2019inverse}, \cite{borowiec2017uncertainty} \cite{wu2018inverse22}).

The void fractions are measured at four elevations along the test section, three of them are measured by X-ray densitometer, and the upper one is measured by a X-ray CT scanner. The void fraction data measured from the lower to upper location are referred to as `Void Fraction 1' to `Void Fraction 4', respectively. BFBT benchmark includes various assembly types and in each assembly type, experiments were conducted at various boundary conditions and void fraction data were measured. 

\begin{figure}[!htbp]
    \centering
    \includegraphics[width=0.75\textwidth]{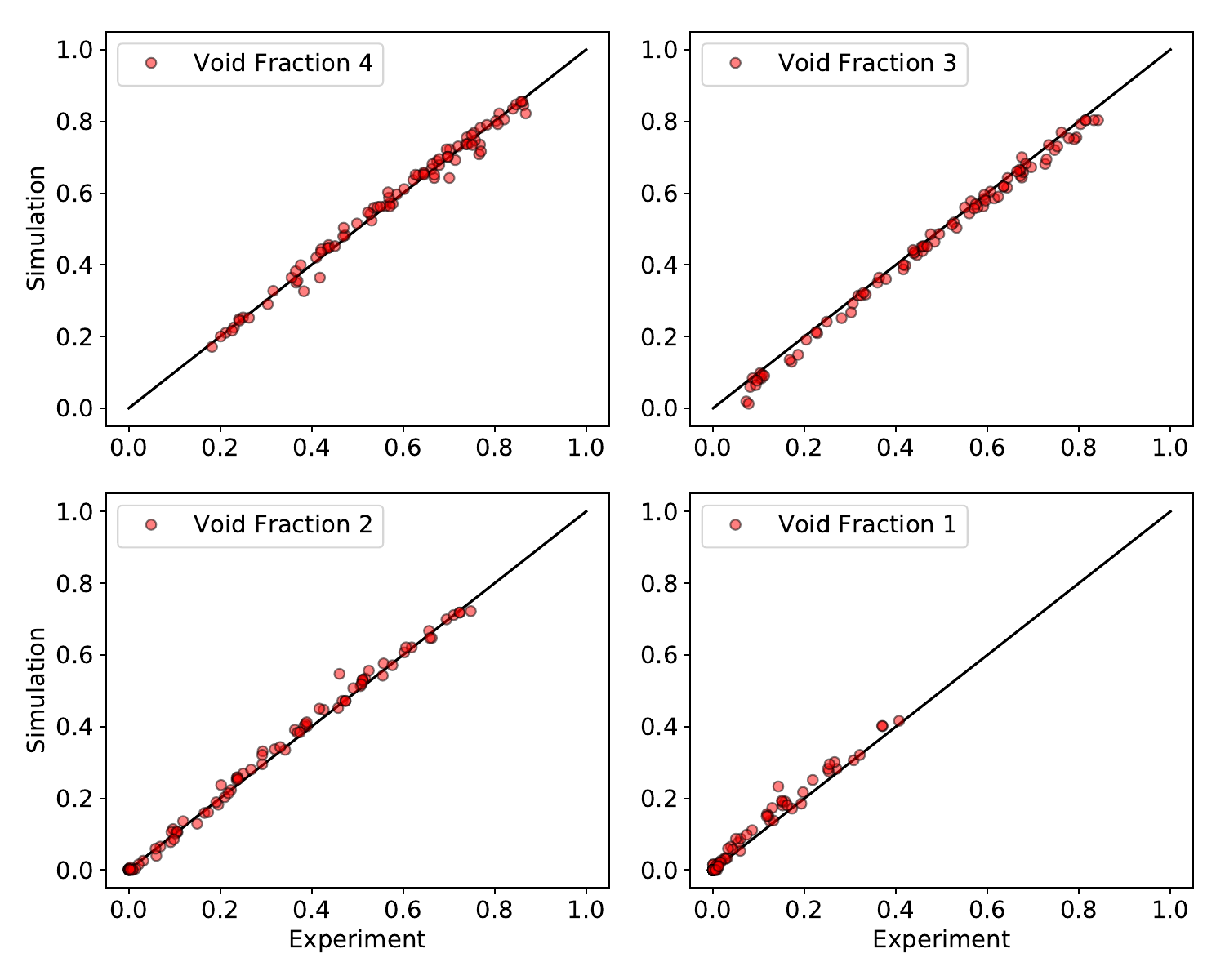}
    \caption{Comparison: predicted void fraction by TRACE versus experimental measurements from Assembly 4 in BFBT benchmark.}
    \label{fig:bfbt-compare}
\end{figure}

The BFBT benchmark consists of 5 assembly types in all. The Assembly-4 is chosen in the current work. It consists of 86 experimental cases under various boundary conditions (flow rate, inlet temperature, power, and inlet pressure). The TRACE model is constructed according to the experimental geometry and boundary conditions. A comparison between the simulated void fraction results and experimental measurements are shown in Figure \ref{fig:bfbt-compare}. The experimental data on `Void Fraction 1', `Void Fraction 2', and `Void Fraction 3' by densitometers are corrected according to \cite{gluck2008validation}. We can see that TRACE model results agree well with experimental data and no obvious model discrepancy exists.

\subsection{Selection of input parameters}

TRACE provides the option to adjust 36 physical model parameters (PMPs) by a multiplicative factor, however, not all of those parameters will be active in the BFBT bundle assembly model because some parameters involve phenomena that do not occur in BFBT benchmark, e.g. reflood. A simple perturbation method is firstly used to perturb each parameter in the range of $(0,5)$ while fixing other parameters. 50 uniform samples in that range are used to test the effect of this parameter on the simulated void fraction data. The resulting output variance is calculated for each parameter. The results show that most of the variances are 0 or very close to 0. Finally, eight parameters with variances larger than $10^{-3}$ are selected and shown in Table \ref{tab:parameter_describ}.

\begin{table}[h!]
    \centering
    \caption{List of 8 selected PMPs in TRACE}
    \begin{tabular}[c]{m{5em}  c}
    \hline
    \textbf{Parameter}     &  \textbf{coefficient Definition} \\
    \hline
    P1000 & Liquid to interface bubbly-slug heat transfer\\
    P1002 & Liquid to interface transition heat transfer  \\
    P1008     & Single phase liquid to wall heat transfer  \\
    P1012     & Subcooled boiling heat transfer    \\
    P1022     & Wall drag  \\
    P1028     & Interfacial drag (bubbly/slug rod bundle-bestion) \\
    P1029 & Interfacial drag (bubbly/slug Vessel) \\
    P1030 & Interfacial drag (annular/mist Vessel) \\
    \hline
    \end{tabular}
    \label{tab:parameter_describ}
\end{table}

\begin{figure}[!h]
    \centering
    \includegraphics[width = 0.8\textwidth]{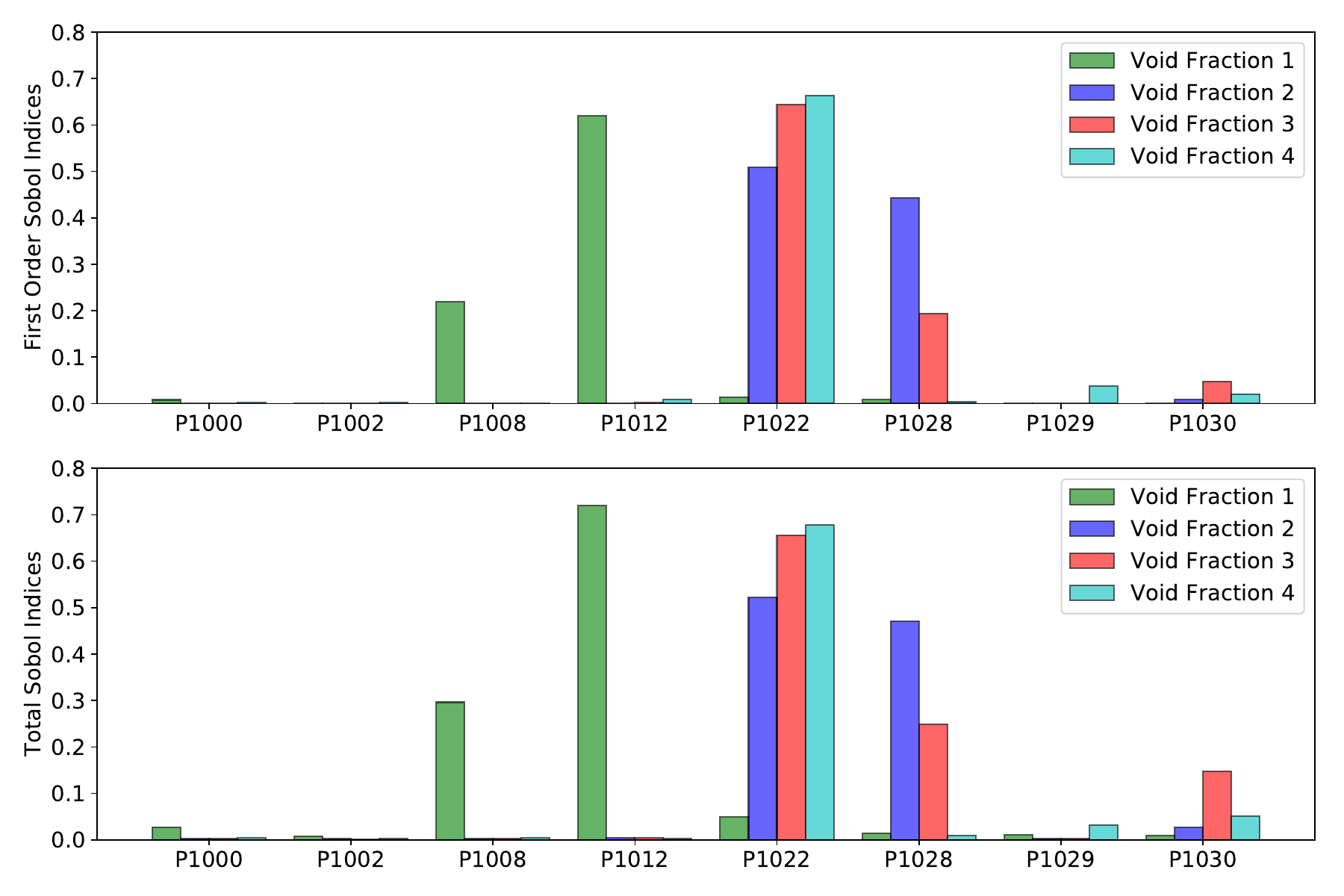}
    \caption{First-order and total Sobol indices.}
    \label{fig:sa-sobol}
\end{figure}

The Sobol indices method is subsequently employed for an further screening, and the first and total Sobol indices are shown in Figure \ref{fig:sa-sobol}. The different colors represent the corresponding indices for a certain measurement location. We can see that four out of eight have more significant influences on the model outputs, so the four parameters P1008, P1012, P1022, and P1028 are selected in the sensitivity analysis step and will be treated as uncertain inputs.

\subsection{Construction of surrogate models}

\begin{figure}[!h]
    \centering
    \includegraphics[width = 0.6\textwidth]{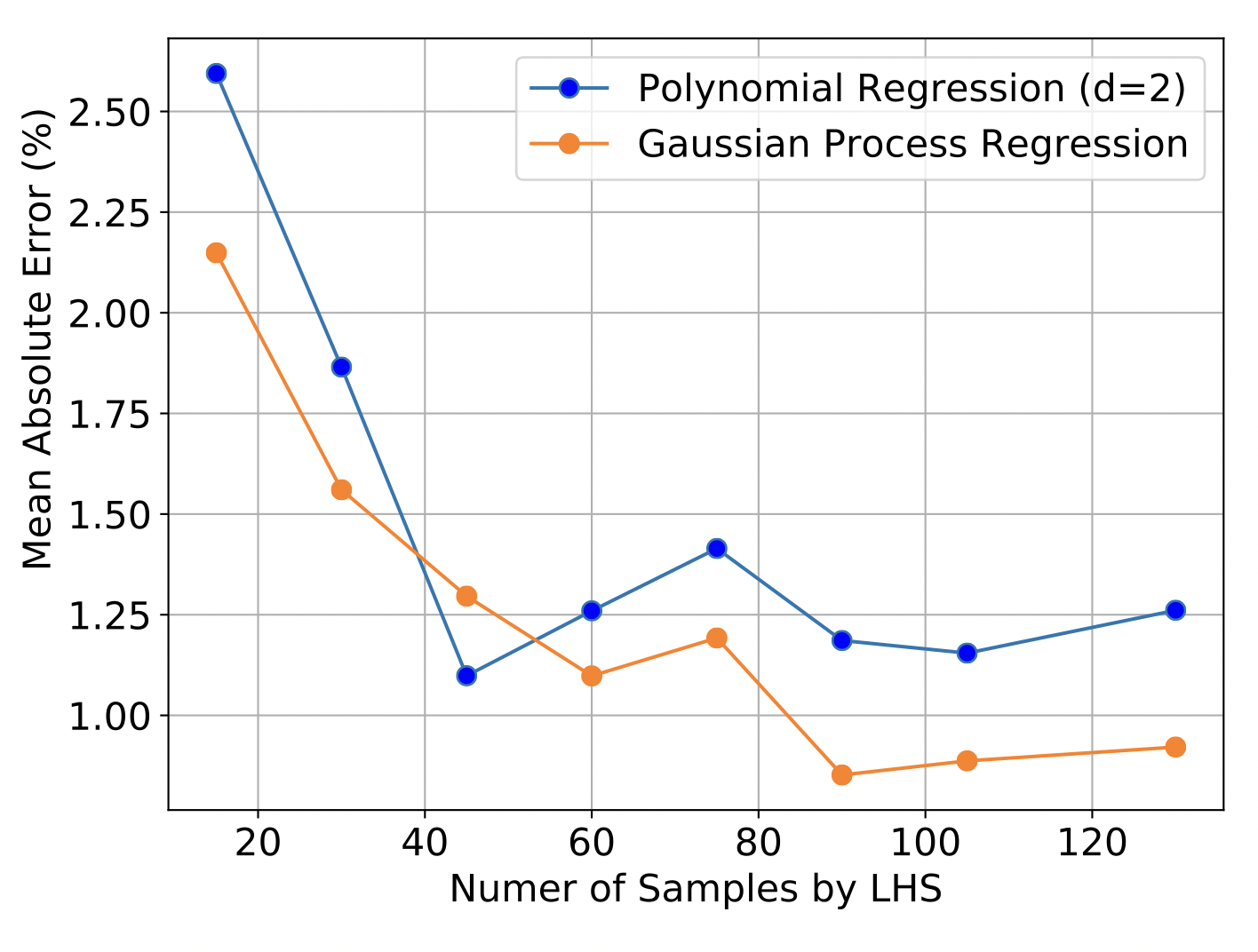}
    \caption{Convergence study to determine the number of necessary LHS samples.}
    \label{fig:surro}
\end{figure}

In this section, we provide the specifics of surrogate models that were introduced earlier. The task is to create a surrogate model with 4 inputs and 4 outputs. In order to determine the number of samples required for an accurate surrogate model, a convergence study is needed. Below we show a convergence study to find the sufficient number of Latin Hypercube Sampling (LHS) samples required by surrogates. Two types of surrogates are considered, GP and PR with a degree of 2. An additional 50 samples are taken as a validation set. Figure \ref{fig:surro} displays the Mean Absolute Error (MAE). The graph indicates that the out-of-sample errors plateau once the sample number surpasses 100. Therefore, in this study, we choose 100 LHS samples to build the surrogate models. For the results presented below, PR model is used because they can provide convenient gradient information. 

In order to determine the range of the input parameter for sampling, an iterative sampling procedure is used in this work. In the posterior sampling phase, it's possible that the posterior distribution of the calibration parameter $\bm \theta$ may exceed the specified prior range. This occurrence is particularly common in the context of certain PMPs, such as the interfacial drag coefficient, where deviations from their nominal value can be large. In these instances, the posterior distribution may be artificially truncated by the upper bound of the initial range. 

To circumvent this issue, we have employed an iterative resampling procedure in \cite{wang2020hierarchical}. If the posterior distribution is truncated by the upper bound, this suggests that the true posterior distribution may exceed the current range. In such a scenario, it is necessary to extend the initial upper bound of this parameter to a higher value. Doing so ensures that the prior range can accommodate the final posterior distribution. This process involves performing new LHS of the original code, the construction of a surrogate model, and the application of MCMC sampling. As the calibration parameters selected in this paper are all multiplicative factors, the lower bound remains 0.

\subsection{Hierarchical Model Structure}

In TRACE, dimensionless multipliers are usually used because it provides a straightforward way to compare those coefficients with the nominal value of $1.0$, and there is no need to calculate their absolute values. Traditional calibration methods implicitly assume that $\bm \theta$ is a global variable for all experimental cases $\bm y^E$. This assumption greatly simplifies the problem being considered, however, it may be problematic in the context of closure models in TH codes. The closure models are derived from Separate Effects Tests and are strongly dependent on boundary conditions and flow regimes, thus the ``true'' parameter $\bm \theta$ may have different distributions (from run to run over the physical experiments), resulting in different posteriors (including the uncertainty information as well as the best-fitting value information of the parameter) sampled by MCMC algorithms.

Ignoring this fact and sampling a single global variable might lead to unrealistic posteriors. An accurate way to treat this problem is to partition the experimental data and simulation outputs according to each correlation equation in a certain range of boundary conditions. However, the work will be tedious and is not meaningful in BEPU applications, because simple input distributions are desired rather than many input distributions dependent on other variables. As a middle ground between the two treatments, we propose to use hierarchical models to account for both similarities and differences among calibration parameters over all considered physical experiments.

The hierarchical model used in this work allows for the calibration parameters $\bm \theta_i$ to vary for each experimental case $i$, reflecting the different boundary conditions. At the same time, we think that those parameters (multipliers) should not display substantial disparities and they can follow a shared distribution, for example, normal distribution. This treatment permits variability for input parameters, which is more realistic and can help reduce the over-fitting issue.

\begin{figure}[!htbp]
    \centering
    \includegraphics[width = 0.7\textwidth]{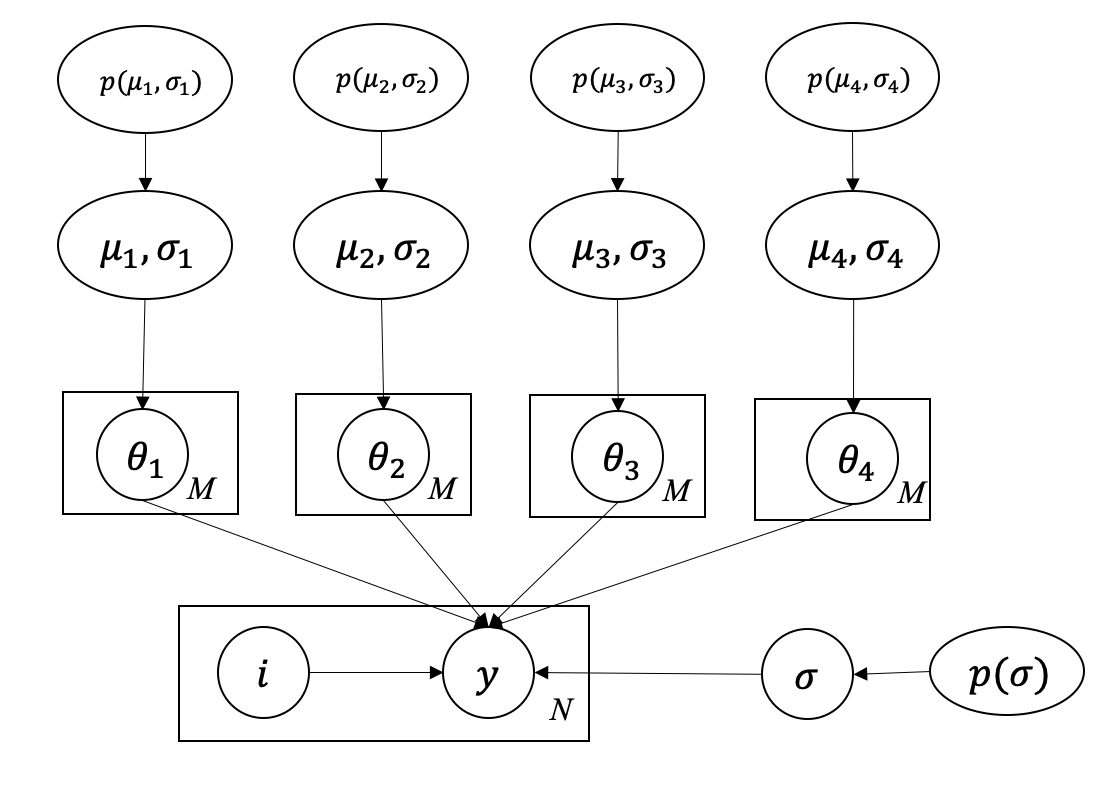}
    \caption{Hierarchical Structure applied to TRACE BFBT application.}
    \label{fig:hierar-trace-psbt}
\end{figure}

In the hierarchical model proposed for this application, we assume that each experimental case forms a `group` or `cluster` since they have different boundary conditions but they have the same assembly type and similar underlying physics. Thus, we can view the 86 experimental cases as forming 86 distinct clusters. Furthermore, each group includes four measurements, leading to the model structure shown in Figure~\ref{fig:hierar-trace-psbt}.

In Figure~\ref{fig:hierar-trace-psbt}, $\theta_i,i=1,2,3,4$ represent our selected four PMPs, `P1008', `P1012', 'P1022', 'P1028', respectively. M is the number of groups and N is the total number of measurement points. In this case $N = 4M$ because there are four measurements in each group. $\sigma$ is measurement error. For each calibration parameter $\theta_i$, since they can be different across groups, so we suppose they are from a common normal distribution:

$$
\theta_i \sim N(\mu_i, \sigma_i)
$$

This approach accounts for possible variations of $\theta_i$ across different experiments that could arise due to potential errors or discrepancies. It is worth noting that some experimental data may exhibit a significant discrepancy for reasons that remain unknown or unaccounted for. Such data points could disproportionately influence the likelihood function, thereby making the posterior distribution highly sensitive to these points. We can view these points as ``outliers''. By employing a hierarchical framework, we can focus on the distribution of $\bm \theta$, making the model robust against outliers. Adding suitable prior information to $\bm \mu$ and $\bm \sigma$ allows us to efficiently perform Bayesian inference for these parameters using the NUTS algorithm.

\subsection{Comparison between the Non-Hierarchical and Hierarchical Models}
\subsubsection{Results of the Non-hierarchical Model}
In order to provide a basis for comparison, we first display the results derived from a non-hierarchical model. Here, only four input parameters are required to estimate uncertainty in the Bayesian model. We use the NUTS sampler to sample the posterior distribution, with the polynomial regression model serving as the foundation. Uniform priors on the interval (0,3) are utilized for all four parameters according to Equation \ref{eqa:23}. For the NUTS sampling algorithm, 100,000 samples are generated and the first 20,000 samples are discarded as burn-in. Multiple chains are used to check the convergence of the Markov chain. It is worth to mention that the non-hierarchical model doesn’t require such many samples, this is only used for the purpose of comparing with hierarchical model.

As we mentioned before, the non-hierarchical model might produce different outcomes when provided with various datasets. Some ``outliers'' may have significant impacts on the results because they may have larger errors thus create more ``weight'' on the likelihood function of the posterior formulation. This effect is evident when we partition the data into two sets.

\begin{figure}[!htbp]
    \centering
    \includegraphics[width = 0.8\textwidth]{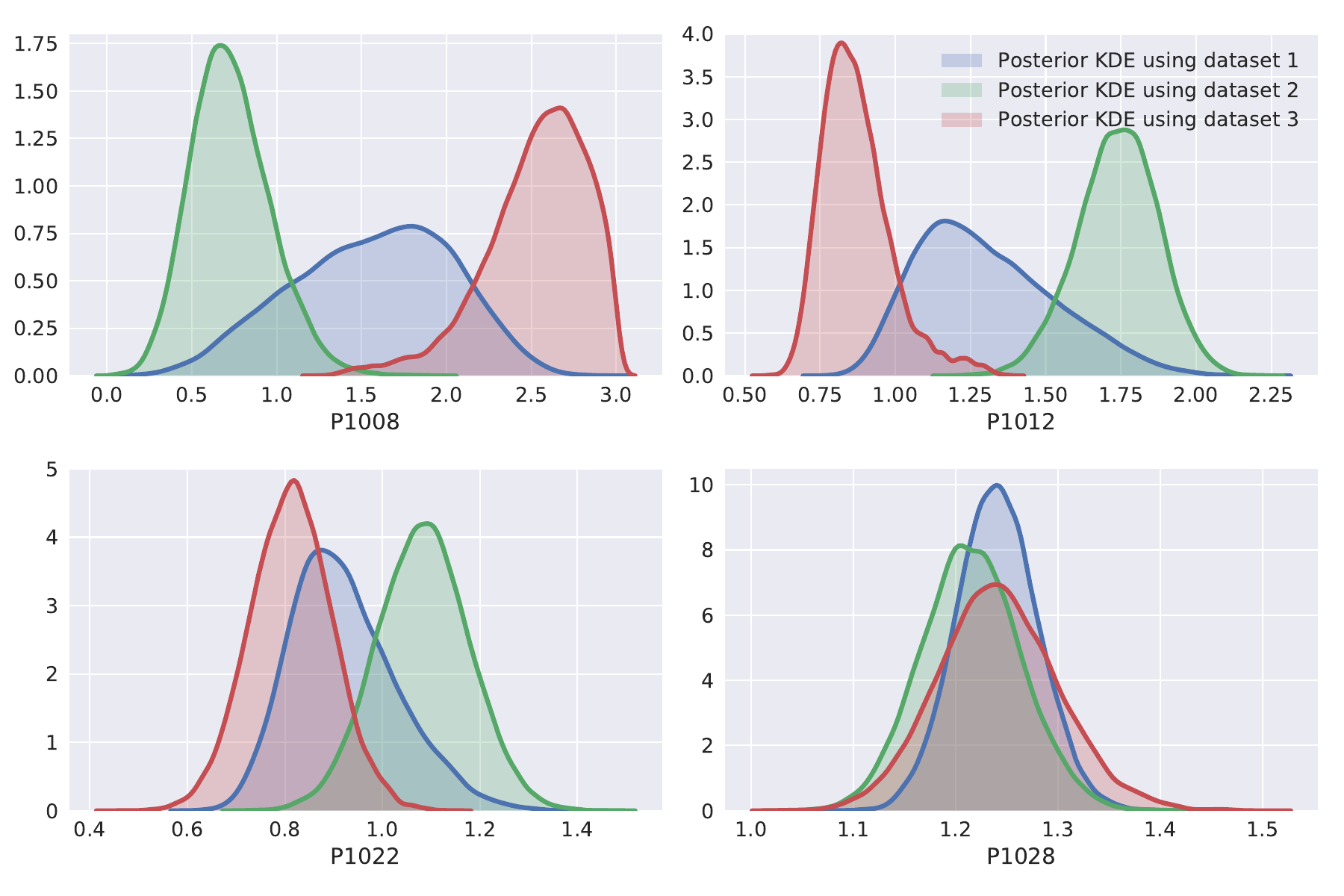}
    \caption{Comparison of posteriors generated by the non-hierarchical model using different datasets (Dataset 1: all data, Dataset 2: randomly selected half of Dataset 1, Dataset 3: remaining half of Dataset 1).}
    \label{fig:fp-com}
\end{figure}

Figure \ref{fig:fp-com} presents the outcomes obtained using a non-hierarchical model with varied datasets. Dataset 1 includes all the data, while Dataset 2 consists of a randomly selected half of Dataset 1's data. The posterior resulting from Dataset 2 yields the green distribution plot. Dataset 3 contains the remaining data from Dataset 1 and is represented by the red distribution plot. This figure clearly illustrates the significant effect of data selection using the traditional method.

\subsubsection{Results of the Hierarchical Model}

In the hierarchical model, the priors of the parameters are formulated as:
\begin{equation}
    \label{eqa:5-datagenerate}
\begin{aligned}
    \mu_{P1008} & \sim \mathbf{Unif}(0,3)\\
    \sigma_{P1008} & \sim \mathbf{Unif}(0,1)\\
    P1008 & \sim \mathcal{N}(\mu_{P1008},\sigma_{P1008})\\
    \sigma & \sim \mathbf{Unif}(0,1)
\end{aligned}
\end{equation}

Other parameters (P1012, P1022, and P1028) are the same and will not be repeated here. It is worth to mention that the range for $\sigma$ is quite large considering that the void fractions are all lower than $1.0$. 

\begin{figure}[!htbp]
    \centering
    \includegraphics[width = 0.8\textwidth]{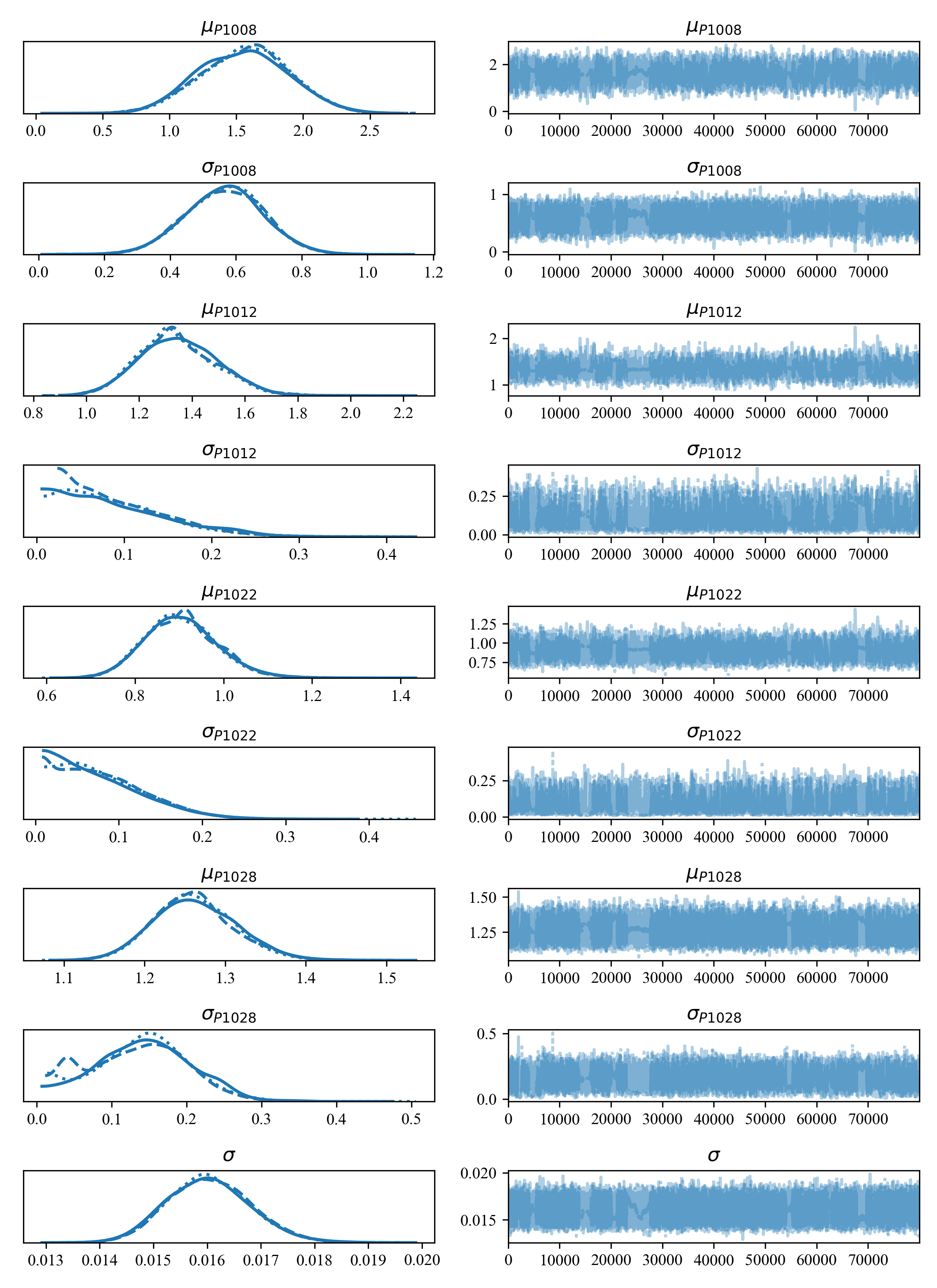}
    \caption{Posterior distributions of all shared parameters by the hierarchical model.}
    \label{fig:hb-1}
\end{figure}

The posteriors and trace plots of all shared parameters ($\mu_{P10**},\sigma_{P10**},\sigma$) calculated by this hierarchical structure using NUTS are shown in Figure \ref{fig:hb-1}. 100,000 steps are used in the MCMC algorithm and the first 20,000 are used as the burn-in period to help convergence. We can see that multiple chains (indicated by different line styles) show similar results and good convergences are achieved. The last parameter in Figure \ref{fig:hb-1} is the total variance term $\Sigma_t$ in the model updating equation. We can see its posterior mean corresponds to a 4\% void fraction error, which is consistent with the reported measurement error in BFBT benchmark.

\begin{figure}[!htbp]
    \centering
    \includegraphics[width = 0.8\textwidth]{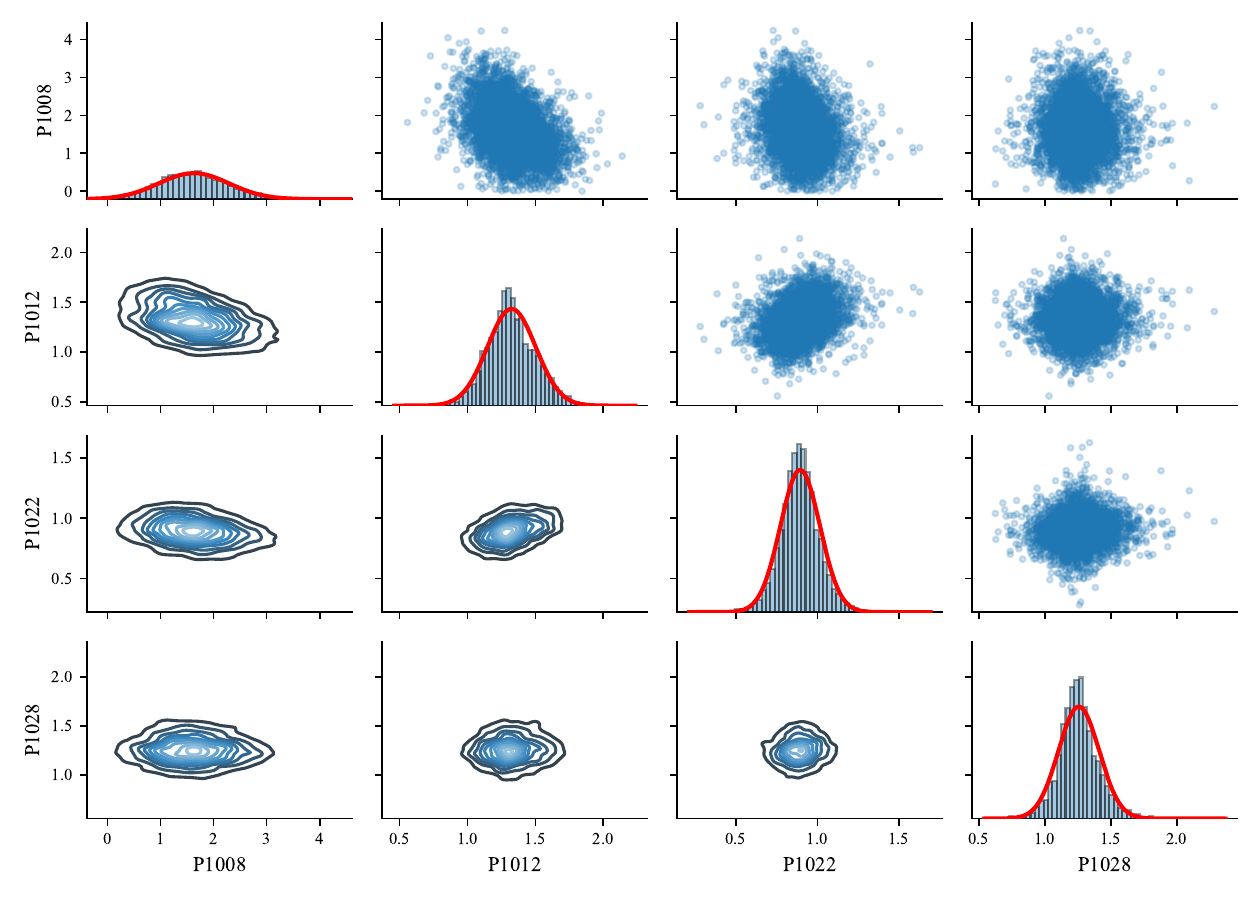}
    \caption{Joint and marginal posterior distributions of PMPs by the hierarchical model.}
    \label{fig:hb-3}
\end{figure}

Now we can use these parameters in Figure \ref{fig:hb-1} to generate samples of the calibration parameters. Note that this is different from the posteriors of cluster-specific calibration parameters because we are not interested in obtaining a distribution for each experimental case. In BEPU applications, we are more interested in a single distribution to describe a parameter. This can be done by the same data generating process shown in Equation \ref{eqa:5-datagenerate}.
For example, recall that the samples of $P1008$ are generated from $\mathcal{N}(\mu_{P1008},\sigma_{P1008})$. $N$ samples are firstly drawn from the posterior of $\mu_{P1008},\sigma_{P1008}$. Then, for each sample, a random normal number is generated where the mean and standard deviation are based on these two parameter. The total of these $N$ numbers will form the posterior of calibration parameters. Take $N=5000$ and the posteriors as well as pair-wise distributions are shown in Figure \ref{fig:hb-3}. The red curves are fitted normal distributions, and the statistics of the fitted distribution for the four parameters are shown in Table \ref{tab:5fitted}. The fitted normal distribution will be very convenient in future UQ or SA. The pair-wise distributions in Figure \ref{fig:hb-3} show positive correlation between P1008 and P1012, and negative correlation between P1012 and P1022. Further studies focusing on the meaning and impacts of input correlations in the calibration process are required in future.

\begin{table}[!htbp]
\centering
\caption{ Fitted distribution for the PMPs.}
\begin{tabular}{c c c c}
\hline
Parameters & Distributions & Dist. Parameter 1 & Dist. Parameter 2 \\
\hline
$P1008$ & Normal & $\mu = 1.63$ & $\sigma =0.66$  \\
$P1012$ & Normal & $\mu = 1.32$  & $\sigma = 0.18$\\
$P1022$ & Normal & $\mu = 0.89$  & $\sigma = 0.12$  \\
$P1028$ & Normal & $\mu = 1.26$  & $\sigma = 0.15 $\\

\hline
\label{tab:5fitted}
\end{tabular}
\end{table}

\begin{figure}[!htbp]
    \centering
    \includegraphics[width = 0.7\textwidth]{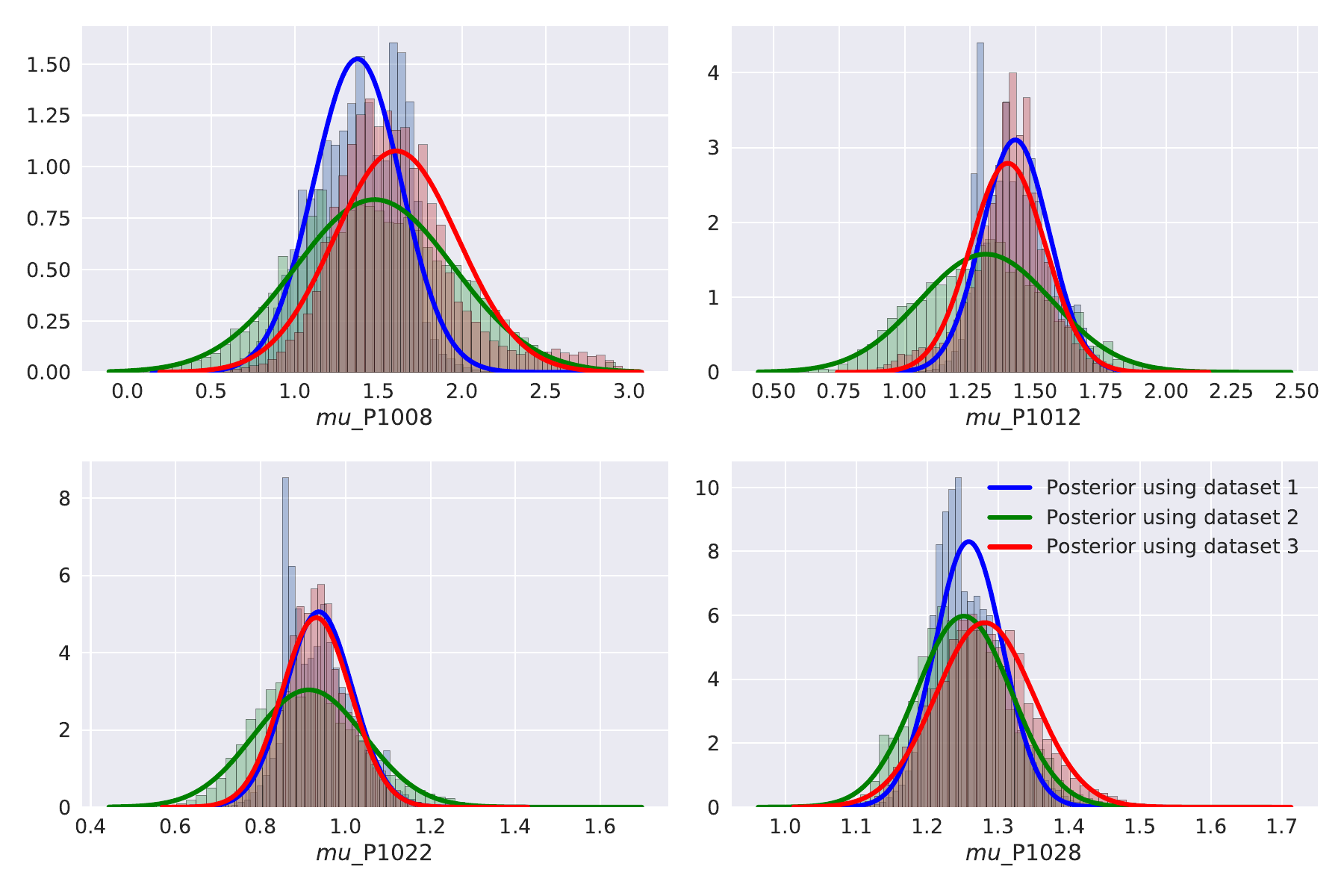}
    \caption{Comparison of posteriors generated by the hierarchical model using different datasets (Dataset 1: all data, Dataset 2: randomly selected half of Dataset 1, Dataset 3: remaining half of Dataset 1).}
    \label{fig:hb-2}
\end{figure}

As a comparison to Figure \ref{fig:fp-com} by the non-hierarchical model, Figure \ref{fig:hb-2}
shows the posteriors using the same 3 datasets described in the previous section. We can see that the hierarchical model is significantly more robust against the impact of data selection. This resilience can be attributed to the fact that the hierarchical model is largely immune to the effects of outliers. If we assume a global variable, every single data point contributes equally to the likelihood. This can make the likelihood particularly sensitive to extreme data. In contrast, the hierarchical model permits variability within calibration parameters, making it less vulnerable to outliers and capable of producing consistent results given sufficient data.

The validity of the obtained posteriors can be confirmed through a process known as a posterior predictive check (PPC). Essentially, this involves conducting a forward Uncertainty Quantification (UQ) process: employing the posteriors as input distributions and propogating these uncertainties through the computational model. This enables us to evaluate the mean and the $95\%$ confidence interval of the model's responses for validation.

\begin{figure}[!htbp]
    \centering
    \includegraphics[width = 0.7\textwidth]{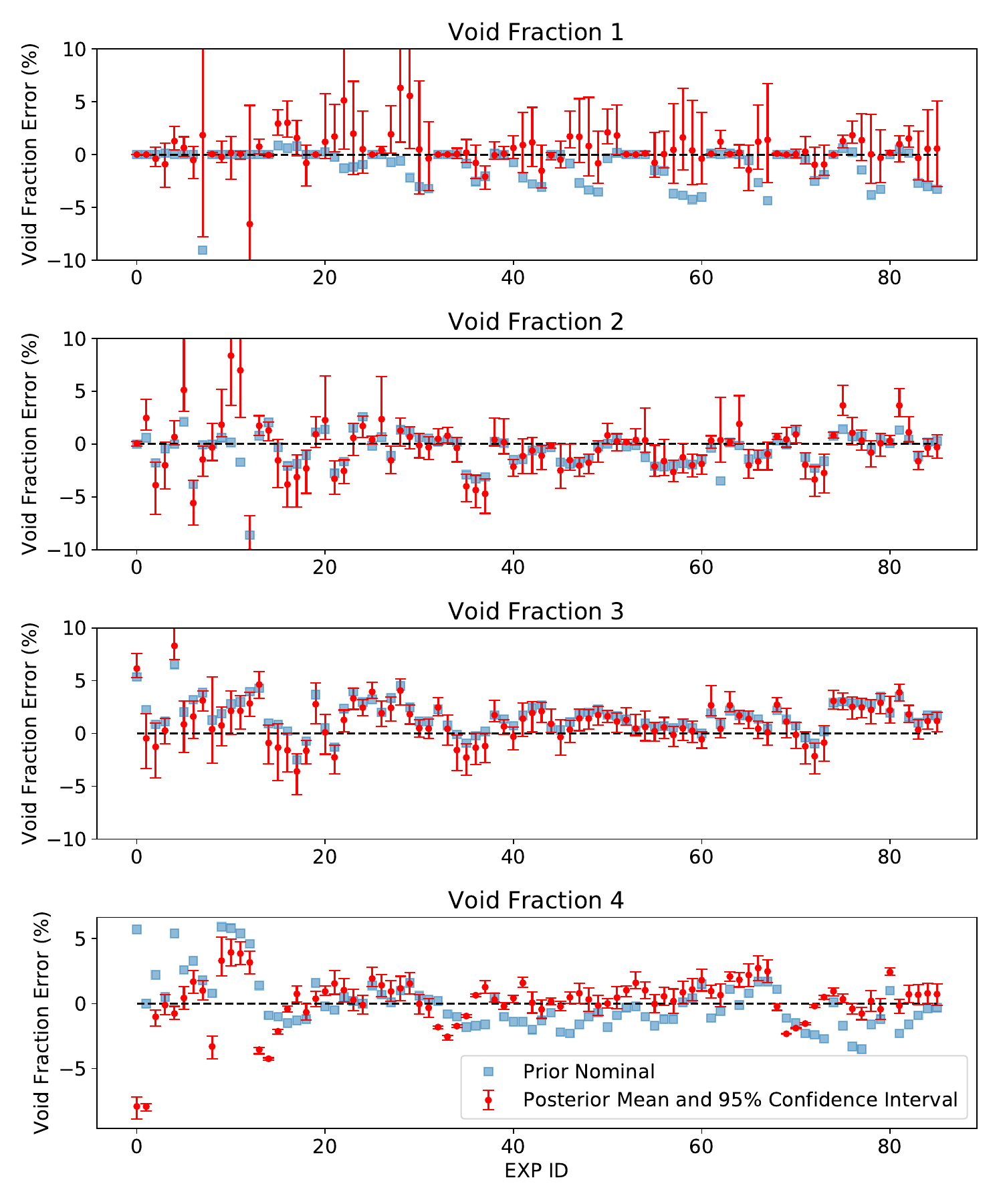}
    \caption{Comparison of TRACE posterior means (red dots) and TRACE output with prior nominal values (blue squares). The error bars represent the $95\%$ confidence interval of the model responses for each case.}
    \label{fig:5vali}
\end{figure}

Figure \ref{fig:5vali} shows the validation of the posteriors. The input samples were drawn from the fitted normal distribution shown in Table \ref{tab:5fitted}. The $95\%$ confidence interval can be seen as the uncertainties in the model responses caused by the parametric uncertainty. We can see that the Void Fraction 1 has relatively larger uncertainty than other locations, and experimental cases with smaller IDs generally have larger uncertainties than those with larger experimental IDs. This can be related to the fact that the inlet pressure increases with the experimental ID, and it is widely known that the TH simulations under low pressures can be problematic.

The posteriors' validity is further confirmed in Figure \ref{fig:5vali} by the improved agreement between simulation results and experimental data. This improvement is readily observable from the Void Fraction 1 and Void Fraction 4. The discrepancies between the experiment and two models (the original model and the calibrated model) are reported in Table \ref{tab:5error}. The original model employs the nominal value of calibration parameters (1.0), and the calibrated model uses the posterior samples of calibration parameters. As shown in Table \ref{tab:5error}, the calibrated model shows a better alignment than the original model for 3 out of 4 measurement locations.

\begin{table}[!htbp]
    \centering
    \caption{Comparison of prediction errors between original model and calibrated model.}
    \begin{tabular}{|c|>{\centering\arraybackslash}m{7em}|>{\centering\arraybackslash}m{4em}|>{\centering\arraybackslash}m{4em}|}
    \hline
         Error types & QoIs & Original model & Calibrated model   \\ 
         \hline
        \multirow{4}{*}{Mean Squared Error [\%]} & Void Fraction 1 & 3.93 & 2.67 \\
                                            & Void Fraction 2 & 2.88 & 3.51 \\
                                            & Void Fraction 3 & 5.09 & 4.77 \\
                                            & Void Fraction 4 & 4.03 & 3.77 \\
        \hline
        \multirow{4}{*}{Mean Absolute Error [\%]} & Void Fraction 1 & 1.18 & 0.99 \\
                                             & Void Fraction 2 & 1.19 & 1.60 \\
                                             & Void Fraction 3 & 1.84 & 1.71 \\
                                             & Void Fraction 4 & 1.51 & 1.31 \\
                                            \hline
    \end{tabular}
    \label{tab:5error}
\end{table}
\section{Conclusions}
\label{sec5}

This work aims at developing a systematic approach for quantifying the input parametric uncertainties in nuclear TH codes to address the ``lack of input uncertainty information'' issue using a novel hierarchical Bayesian model. The traditional single-level IUQ approaches are often constrained to relatively small datasets and the calculated posterior distributions are applicable only for the selected experimental cases and may vary when different datasets are used. Outliers also have relative large impacts on the single-level IUQ approach. The proposed hierarchical model considers the variability of calibration parameters and assumes that each calibration parameter is generated from a common population distribution. Using the hierarchical model can help make the model robust against outliers and avoid over-fitting. 

The hierarchical model in IUQ is beneficial for nuclear TH applications because the PMPs exhibit group-specific behaviours that require hierarchical structures. The proposed approach is applied to a BFBT benchmark dataset using the TRACE code. In order to deal with the high dimension calibration problem under the hierarchical model, NUTS is used to more efficiently sample the posterior distribution. PPC is used to verify that hierarchical models deliver better alignment with experimental data compared to the original model. Additionally, the IUQ results obtained from the hierarchical model were found to be more robust against outliers and can provide consistent results irrespective of data selection. The hierarchical model also shows a promising approach for generalizing to larger databases with broad ranges of experimental conditions and different geometric setups.

In the next steps, we intend to expand the capacity of our current IUQ framework to accommodate a larger pool of observations. Nevertheless, the computational demands posed by the hierarchical model and the MCMC algorithms represent a significant bottleneck, particularly when the hierarchical structure involves hundreds of groups. More efficient MCMC algorithms such as Variational Inference should be explored and validated. In addition, model discrepancy always poses challenges for IUQ of computer code and Modular Bayesian Approach can be useful when model discrepancy exists. In future works, we need to consider a combination of the Modular Bayesian approach and hierarchical model to in order to apply the IUQ framework for broader applications when model discrepancy exists.



\bibliographystyle{elsarticle-harv} 
\bibliography{main}

\begin{thebibliography}{52}
\expandafter\ifx\csname natexlab\endcsname\relax\def\natexlab#1{#1}\fi
\providecommand{\url}[1]{\texttt{#1}}
\providecommand{\href}[2]{#2}
\providecommand{\path}[1]{#1}
\providecommand{\DOIprefix}{doi:}
\providecommand{\ArXivprefix}{arXiv:}
\providecommand{\URLprefix}{URL: }
\providecommand{\Pubmedprefix}{pmid:}
\providecommand{\doi}[1]{\href{http://dx.doi.org/#1}{\path{#1}}}
\providecommand{\Pubmed}[1]{\href{pmid:#1}{\path{#1}}}
\providecommand{\bibinfo}[2]{#2}
\ifx\xfnm\relax \def\xfnm[#1]{\unskip,\space#1}\fi
\bibitem[{Aly et~al.(2019)Aly, Casagranda, Pastore and Brown}]{aly2019variance}
\bibinfo{author}{Aly, Z.}, \bibinfo{author}{Casagranda, A.},
  \bibinfo{author}{Pastore, G.}, \bibinfo{author}{Brown, N.R.},
  \bibinfo{year}{2019}.
\newblock \bibinfo{title}{Variance-based sensitivity analysis applied to the
  hydrogen migration and redistribution model in bison. part ii: Uncertainty
  quantification and optimization}.
\newblock \bibinfo{journal}{Journal of Nuclear Materials}
  \bibinfo{volume}{523}, \bibinfo{pages}{478--489}.
\bibitem[{Amri and Gulliford(2013)}]{amri2013overview}
\bibinfo{author}{Amri, A.}, \bibinfo{author}{Gulliford, J.},
  \bibinfo{year}{2013}.
\newblock \bibinfo{title}{Overview of OECD/NEA BEPU Programmes}.
\newblock \bibinfo{type}{Technical Report}.
\bibitem[{Andrieu and Thoms(2008)}]{andrieu2008tutorial}
\bibinfo{author}{Andrieu, C.}, \bibinfo{author}{Thoms, J.},
  \bibinfo{year}{2008}.
\newblock \bibinfo{title}{A tutorial on adaptive mcmc}.
\newblock \bibinfo{journal}{Statistics and computing} \bibinfo{volume}{18},
  \bibinfo{pages}{343--373}.
\bibitem[{Baccou et~al.(2020)Baccou, Zhang, Fillion, Damblin, Petruzzi,
  Mendiz{\'a}bal, Reventos, Skorek, Couplet, Iooss et~al.}]{baccou2020sapium}
\bibinfo{author}{Baccou, J.}, \bibinfo{author}{Zhang, J.},
  \bibinfo{author}{Fillion, P.}, \bibinfo{author}{Damblin, G.},
  \bibinfo{author}{Petruzzi, A.}, \bibinfo{author}{Mendiz{\'a}bal, R.},
  \bibinfo{author}{Reventos, F.}, \bibinfo{author}{Skorek, T.},
  \bibinfo{author}{Couplet, M.}, \bibinfo{author}{Iooss, B.}, et~al.,
  \bibinfo{year}{2020}.
\newblock \bibinfo{title}{Sapium: A generic framework for a practical and
  transparent quantification of thermal-hydraulic code model input
  uncertainty}.
\newblock \bibinfo{journal}{Nuclear Science and Engineering}
  \bibinfo{volume}{194}, \bibinfo{pages}{721--736}.
\bibitem[{Bajorek et~al.(2008)}]{bajorek2008trace}
\bibinfo{author}{Bajorek, S.}, et~al., \bibinfo{year}{2008}.
\newblock \bibinfo{title}{Trace v5. 0 theory manual, field equations, solution
  methods and physical models}.
\newblock \bibinfo{journal}{United States Nuclear Regulatory Commission} .
\bibitem[{Barth(2011)}]{barth2011brief}
\bibinfo{author}{Barth, T.}, \bibinfo{year}{2011}.
\newblock \bibinfo{title}{A brief overview of uncertainty quantification and
  error estimation in numerical simulation}.
\newblock \bibinfo{journal}{NASA Ames Research Center, NASA Report} .
\bibitem[{Borowiec et~al.(2017)Borowiec, Wang, Kozlowski and
  Brooks}]{borowiec2017uncertainty}
\bibinfo{author}{Borowiec, K.}, \bibinfo{author}{Wang, C.},
  \bibinfo{author}{Kozlowski, T.}, \bibinfo{author}{Brooks, C.S.},
  \bibinfo{year}{2017}.
\newblock \bibinfo{title}{Uncertainty quantification for steady-state psbt
  benchmark using surrogate models}.
\newblock \bibinfo{journal}{Transactions of the American Nuclear Society}
  \bibinfo{volume}{117}, \bibinfo{pages}{119--122}.
\bibitem[{Chen(2020)}]{chen2020some}
\bibinfo{author}{Chen, S.}, \bibinfo{year}{2020}.
\newblock \bibinfo{title}{Some recent advances in design of bayesian binomial
  reliability demonstration tests}.
\newblock \bibinfo{publisher}{University of South Florida}.
\bibitem[{Chen et~al.(2019)Chen, Kong, Sun, Meng and Li}]{chen2019claims}
\bibinfo{author}{Chen, S.}, \bibinfo{author}{Kong, N.}, \bibinfo{author}{Sun,
  X.}, \bibinfo{author}{Meng, H.}, \bibinfo{author}{Li, M.},
  \bibinfo{year}{2019}.
\newblock \bibinfo{title}{Claims data-driven modeling of hospital
  time-to-readmission risk with latent heterogeneity}.
\newblock \bibinfo{journal}{Health care management science}
  \bibinfo{volume}{22}, \bibinfo{pages}{156--179}.
\bibitem[{Chen et~al.(2017)Chen, Lu and Li}]{chen2017multi}
\bibinfo{author}{Chen, S.}, \bibinfo{author}{Lu, L.}, \bibinfo{author}{Li, M.},
  \bibinfo{year}{2017}.
\newblock \bibinfo{title}{Multi-state reliability demonstration tests}.
\newblock \bibinfo{journal}{Quality Engineering} \bibinfo{volume}{29},
  \bibinfo{pages}{431--445}.
\bibitem[{Chen et~al.(2018)Chen, Lu, Xiang, Lu and Li}]{chen2018data}
\bibinfo{author}{Chen, S.}, \bibinfo{author}{Lu, L.}, \bibinfo{author}{Xiang,
  Y.}, \bibinfo{author}{Lu, Q.}, \bibinfo{author}{Li, M.},
  \bibinfo{year}{2018}.
\newblock \bibinfo{title}{A data heterogeneity modeling and quantification
  approach for field pre-assessment of chloride-induced corrosion in aging
  infrastructures}.
\newblock \bibinfo{journal}{Reliability Engineering \& System Safety}
  \bibinfo{volume}{171}, \bibinfo{pages}{123--135}.
\bibitem[{Chen et~al.(2020)Chen, Lu, Zhang and Li}]{chen2020optimal}
\bibinfo{author}{Chen, S.}, \bibinfo{author}{Lu, L.}, \bibinfo{author}{Zhang,
  Q.}, \bibinfo{author}{Li, M.}, \bibinfo{year}{2020}.
\newblock \bibinfo{title}{Optimal binomial reliability demonstration tests
  design under acceptance decision uncertainty}.
\newblock \bibinfo{journal}{Quality Engineering} \bibinfo{volume}{32},
  \bibinfo{pages}{492--508}.
\bibitem[{Chen et~al.(2023)Chen, Wu, Hovakimyan and Yao}]{chen2023recontab}
\bibinfo{author}{Chen, S.}, \bibinfo{author}{Wu, J.},
  \bibinfo{author}{Hovakimyan, N.}, \bibinfo{author}{Yao, H.},
  \bibinfo{year}{2023}.
\newblock \bibinfo{title}{Recontab: Regularized contrastive representation
  learning for tabular data}.
\newblock \bibinfo{journal}{arXiv preprint arXiv:2310.18541} .
\bibitem[{Gelman et~al.(2013)Gelman, Carlin, Stern, Dunson, Vehtari and
  Rubin}]{gelman2013bayesian}
\bibinfo{author}{Gelman, A.}, \bibinfo{author}{Carlin, J.B.},
  \bibinfo{author}{Stern, H.S.}, \bibinfo{author}{Dunson, D.B.},
  \bibinfo{author}{Vehtari, A.}, \bibinfo{author}{Rubin, D.B.},
  \bibinfo{year}{2013}.
\newblock \bibinfo{title}{Bayesian data analysis}.
\newblock \bibinfo{publisher}{CRC press}.
\bibitem[{Glaeser et~al.(2011)Glaeser, Bazin, Baccou, Chojnacki and
  Destercke}]{glaeser2011bemuse}
\bibinfo{author}{Glaeser, H.}, \bibinfo{author}{Bazin, P.},
  \bibinfo{author}{Baccou, J.}, \bibinfo{author}{Chojnacki, E.},
  \bibinfo{author}{Destercke, S.}, \bibinfo{year}{2011}.
\newblock \bibinfo{title}{Bemuse phase vi report, status report on the area,
  classification of the methods, conclusions and recommendations}.
\newblock \bibinfo{journal}{Nuclear Energy Agency Committee on the Safety of
  Nuclear Installations} .
\bibitem[{Gl{\"u}ck(2008)}]{gluck2008validation}
\bibinfo{author}{Gl{\"u}ck, M.}, \bibinfo{year}{2008}.
\newblock \bibinfo{title}{Validation of the sub-channel code f-cobra-tf: Part
  ii. recalculation of void measurements}.
\newblock \bibinfo{journal}{Nuclear Engineering and Design}
  \bibinfo{volume}{238}, \bibinfo{pages}{2317--2327}.
\bibitem[{Hoffman and Gelman(2014)}]{hoffman2014no}
\bibinfo{author}{Hoffman, M.D.}, \bibinfo{author}{Gelman, A.},
  \bibinfo{year}{2014}.
\newblock \bibinfo{title}{The no-u-turn sampler: adaptively setting path
  lengths in hamiltonian monte carlo.}
\newblock \bibinfo{journal}{Journal of Machine Learning Research}
  \bibinfo{volume}{15}, \bibinfo{pages}{1593--1623}.
\bibitem[{Kass et~al.(1998)Kass, Carlin, Gelman and Neal}]{kass1998markov}
\bibinfo{author}{Kass, R.E.}, \bibinfo{author}{Carlin, B.P.},
  \bibinfo{author}{Gelman, A.}, \bibinfo{author}{Neal, R.M.},
  \bibinfo{year}{1998}.
\newblock \bibinfo{title}{Markov chain monte carlo in practice: a roundtable
  discussion}.
\newblock \bibinfo{journal}{The American Statistician} \bibinfo{volume}{52},
  \bibinfo{pages}{93--100}.
\bibitem[{Kennedy and O'Hagan(2001)}]{kennedy2001bayesian}
\bibinfo{author}{Kennedy, M.C.}, \bibinfo{author}{O'Hagan, A.},
  \bibinfo{year}{2001}.
\newblock \bibinfo{title}{Bayesian calibration of computer models}.
\newblock \bibinfo{journal}{Journal of the Royal Statistical Society: Series B
  (Statistical Methodology)} \bibinfo{volume}{63}, \bibinfo{pages}{425--464}.
\bibitem[{Liu and Neville(2023)}]{liu2023stationary}
\bibinfo{author}{Liu, J.}, \bibinfo{author}{Neville, J.}, \bibinfo{year}{2023}.
\newblock \bibinfo{title}{Stationary algorithmic balancing for dynamic email
  re-ranking problem}, in: \bibinfo{booktitle}{Proceedings of the 29th ACM
  SIGKDD Conference on Knowledge Discovery and Data Mining}, pp.
  \bibinfo{pages}{4527--4538}.
\bibitem[{Liu et~al.(2024)Liu, Yang and Neville}]{liu2024cliqueparcel}
\bibinfo{author}{Liu, J.}, \bibinfo{author}{Yang, T.},
  \bibinfo{author}{Neville, J.}, \bibinfo{year}{2024}.
\newblock \bibinfo{title}{Cliqueparcel: An approach for batching llm prompts
  that jointly optimizes efficiency and faithfulness}.
\newblock \bibinfo{journal}{arXiv preprint arXiv:2402.14833} .
\bibitem[{Liu et~al.(2019)Liu, Yuan, Yang, Huang, Zhang and Yu}]{liu2019dapred}
\bibinfo{author}{Liu, J.}, \bibinfo{author}{Yuan, Q.}, \bibinfo{author}{Yang,
  C.}, \bibinfo{author}{Huang, H.}, \bibinfo{author}{Zhang, C.},
  \bibinfo{author}{Yu, P.}, \bibinfo{year}{2019}.
\newblock \bibinfo{title}{Dapred: Dynamic attention location prediction with
  long-short term movement regularity} .
\bibitem[{Liu et~al.(2022)Liu, Hu, Zou and Nunez}]{liu2022sam}
\bibinfo{author}{Liu, Y.}, \bibinfo{author}{Hu, R.}, \bibinfo{author}{Zou, L.},
  \bibinfo{author}{Nunez, D.}, \bibinfo{year}{2022}.
\newblock \bibinfo{title}{Sam-ml: Integrating data-driven closure with nuclear
  system code sam for improved modeling capability}.
\newblock \bibinfo{journal}{Nuclear Engineering and Design}
  \bibinfo{volume}{400}, \bibinfo{pages}{112059}.
\bibitem[{Liu et~al.(2021)Liu, Wang, Sun, Dinh and Hu}]{liu2021uncertainty}
\bibinfo{author}{Liu, Y.}, \bibinfo{author}{Wang, D.}, \bibinfo{author}{Sun,
  X.}, \bibinfo{author}{Dinh, N.}, \bibinfo{author}{Hu, R.},
  \bibinfo{year}{2021}.
\newblock \bibinfo{title}{Uncertainty quantification for multiphase-cfd
  simulations of bubbly flows: a machine learning-based bayesian approach
  supported by high-resolution experiments}.
\newblock \bibinfo{journal}{Reliability Engineering \& System Safety}
  \bibinfo{volume}{212}, \bibinfo{pages}{107636}.
\bibitem[{Mendiz{\'a}bal et~al.(2016)Mendiz{\'a}bal, de~Alfonso, Freixa and
  Revent{\'o}s}]{mendizabal2016oecd}
\bibinfo{author}{Mendiz{\'a}bal, R.}, \bibinfo{author}{de~Alfonso, E.},
  \bibinfo{author}{Freixa, J.}, \bibinfo{author}{Revent{\'o}s, F.},
  \bibinfo{year}{2016}.
\newblock \bibinfo{title}{Oecd/nea premium benchmark final report}.
\bibitem[{Neykov et~al.(2006)Neykov, Aydogan, Hochreiter, Ivanov, Utsuno,
  Kasahara, Sartori and Martin}]{neykov2006nupec}
\bibinfo{author}{Neykov, B.}, \bibinfo{author}{Aydogan, F.},
  \bibinfo{author}{Hochreiter, L.}, \bibinfo{author}{Ivanov, K.},
  \bibinfo{author}{Utsuno, H.}, \bibinfo{author}{Kasahara, F.},
  \bibinfo{author}{Sartori, E.}, \bibinfo{author}{Martin, M.},
  \bibinfo{year}{2006}.
\newblock \bibinfo{title}{Nupec bwr full-size fine-mesh bundle test (bfbt)
  benchmark}.
\newblock \bibinfo{journal}{OECD Papers} \bibinfo{volume}{6},
  \bibinfo{pages}{1--132}.
\bibitem[{Perez et~al.(2011)Perez, Reventos, Batet, Guba, T{\'o}th, Mieusset,
  Bazin, De~Cr{\'e}cy, Borisov, Skorek et~al.}]{perez2011uncertainty}
\bibinfo{author}{Perez, M.}, \bibinfo{author}{Reventos, F.},
  \bibinfo{author}{Batet, L.}, \bibinfo{author}{Guba, A.},
  \bibinfo{author}{T{\'o}th, I.}, \bibinfo{author}{Mieusset, T.},
  \bibinfo{author}{Bazin, P.}, \bibinfo{author}{De~Cr{\'e}cy, A.},
  \bibinfo{author}{Borisov, S.}, \bibinfo{author}{Skorek, T.}, et~al.,
  \bibinfo{year}{2011}.
\newblock \bibinfo{title}{Uncertainty and sensitivity analysis of a lbloca in a
  pwr nuclear power plant: Results of the phase v of the bemuse programme}.
\newblock \bibinfo{journal}{Nuclear Engineering and Design}
  \bibinfo{volume}{241}, \bibinfo{pages}{4206--4222}.
\bibitem[{Rasmussen(2004)}]{rasmussen2004gaussian}
\bibinfo{author}{Rasmussen, C.E.}, \bibinfo{year}{2004}.
\newblock \bibinfo{title}{Gaussian processes in machine learning}, in:
  \bibinfo{booktitle}{Advanced lectures on machine learning}.
  \bibinfo{publisher}{Springer}, pp. \bibinfo{pages}{63--71}.
\bibitem[{Saltelli(2002)}]{saltelli2002making}
\bibinfo{author}{Saltelli, A.}, \bibinfo{year}{2002}.
\newblock \bibinfo{title}{Making best use of model evaluations to compute
  sensitivity indices}.
\newblock \bibinfo{journal}{Computer physics communications}
  \bibinfo{volume}{145}, \bibinfo{pages}{280--297}.
\bibitem[{Saltelli et~al.(2010)Saltelli, Annoni, Azzini, Campolongo, Ratto and
  Tarantola}]{saltelli2010variance}
\bibinfo{author}{Saltelli, A.}, \bibinfo{author}{Annoni, P.},
  \bibinfo{author}{Azzini, I.}, \bibinfo{author}{Campolongo, F.},
  \bibinfo{author}{Ratto, M.}, \bibinfo{author}{Tarantola, S.},
  \bibinfo{year}{2010}.
\newblock \bibinfo{title}{Variance based sensitivity analysis of model output.
  design and estimator for the total sensitivity index}.
\newblock \bibinfo{journal}{Computer Physics Communications}
  \bibinfo{volume}{181}, \bibinfo{pages}{259--270}.
\bibitem[{Salvatier et~al.(2016)Salvatier, Wiecki and
  Fonnesbeck}]{salvatier2016probabilistic}
\bibinfo{author}{Salvatier, J.}, \bibinfo{author}{Wiecki, T.V.},
  \bibinfo{author}{Fonnesbeck, C.}, \bibinfo{year}{2016}.
\newblock \bibinfo{title}{Probabilistic programming in python using pymc3}.
\newblock \bibinfo{journal}{PeerJ Computer Science} \bibinfo{volume}{2},
  \bibinfo{pages}{e55}.
\bibitem[{Skorek and Crecy(2013)}]{skorek2013premium}
\bibinfo{author}{Skorek, T.}, \bibinfo{author}{Crecy, A.d.},
  \bibinfo{year}{2013}.
\newblock \bibinfo{title}{PREMIUM-Benchmark on the quantification of the
  uncertainty of the physical models in the system thermal-hydraulic codes}.
\newblock \bibinfo{type}{Technical Report}.
\bibitem[{Smith(2013)}]{smith2013uncertainty}
\bibinfo{author}{Smith, R.C.}, \bibinfo{year}{2013}.
\newblock \bibinfo{title}{Uncertainty quantification: theory, implementation,
  and applications}. volume~\bibinfo{volume}{12}.
\newblock \bibinfo{publisher}{Siam}.
\bibitem[{Tarantola(2005)}]{tarantola2005inverse}
\bibinfo{author}{Tarantola, A.}, \bibinfo{year}{2005}.
\newblock \bibinfo{title}{Inverse problem theory and methods for model
  parameter estimation}. volume~\bibinfo{volume}{89}.
\newblock \bibinfo{publisher}{siam}.
\bibitem[{Wang et~al.(2024)Wang, Lu, Chen and Li}]{wang2024optimal}
\bibinfo{author}{Wang, B.}, \bibinfo{author}{Lu, L.}, \bibinfo{author}{Chen,
  S.}, \bibinfo{author}{Li, M.}, \bibinfo{year}{2024}.
\newblock \bibinfo{title}{Optimal test design for reliability demonstration
  under multi-stage acceptance uncertainties}.
\newblock \bibinfo{journal}{Quality Engineering} \bibinfo{volume}{36},
  \bibinfo{pages}{91--104}.
\bibitem[{Wang(2020)}]{wang2020hierarchical}
\bibinfo{author}{Wang, C.}, \bibinfo{year}{2020}.
\newblock \bibinfo{title}{A hierarchical Bayesian calibration framework for
  quantifying input uncertainties in thermal-hydraulics simulation models}.
\newblock Ph.D. thesis.
\bibitem[{Wang et~al.(2018a)Wang, Wu, Borowiec and Kozlowski}]{wang2018ans}
\bibinfo{author}{Wang, C.}, \bibinfo{author}{Wu, X.},
  \bibinfo{author}{Borowiec, K.}, \bibinfo{author}{Kozlowski, T.},
  \bibinfo{year}{2018}a.
\newblock \bibinfo{title}{Bayesian calibration and uncertainty quantification
  for trace based on psbt benchmark}.
\newblock \bibinfo{journal}{Transactions of the American Nuclear Society}
  \bibinfo{volume}{118}, \bibinfo{pages}{419--422}.
\bibitem[{Wang et~al.(2017)Wang, Wu and Kozlowski}]{wang2017sensitivity}
\bibinfo{author}{Wang, C.}, \bibinfo{author}{Wu, X.},
  \bibinfo{author}{Kozlowski, T.}, \bibinfo{year}{2017}.
\newblock \bibinfo{title}{Sensitivity and uncertainty analysis of trace
  physical model parameters based on psbt benchmark using gaussian process
  emulator}.
\newblock \bibinfo{journal}{Proc. 17th Int. Topl. Mtg. Nuclear Reactor Thermal
  Hydraulics (NURETH-17)} , \bibinfo{pages}{3--8}.
\bibitem[{Wang et~al.(2018b)Wang, Wu and Kozlowski}]{wangsurrogate}
\bibinfo{author}{Wang, C.}, \bibinfo{author}{Wu, X.},
  \bibinfo{author}{Kozlowski, T.}, \bibinfo{year}{2018}b.
\newblock \bibinfo{title}{Surrogate-based bayesian calibration of
  thermal-hydraulics models based on psbt time-dependent benchmark data}, in:
  \bibinfo{booktitle}{Proc. ANS Best Estimate Plus Uncertainty International
  Conference, Real Collegio, Lucca, Italy}.
\bibitem[{Wang et~al.(2019a)Wang, Wu and Kozlowski}]{wang2019gaussian}
\bibinfo{author}{Wang, C.}, \bibinfo{author}{Wu, X.},
  \bibinfo{author}{Kozlowski, T.}, \bibinfo{year}{2019}a.
\newblock \bibinfo{title}{Gaussian process--based inverse uncertainty
  quantification for trace physical model parameters using steady-state psbt
  benchmark}.
\newblock \bibinfo{journal}{Nuclear Science and Engineering}
  \bibinfo{volume}{193}, \bibinfo{pages}{100--114}.
\bibitem[{Wang et~al.(2019b)Wang, Wu and Kozlowski}]{wang2019inverse}
\bibinfo{author}{Wang, C.}, \bibinfo{author}{Wu, X.},
  \bibinfo{author}{Kozlowski, T.}, \bibinfo{year}{2019}b.
\newblock \bibinfo{title}{Inverse uncertainty quantification by hierarchical
  bayesian inference for trace physical model parameters based on bfbt
  benchmark}.
\newblock \bibinfo{journal}{Proceedings of NURETH-2019, Portland, Oregon, USA}
  .
\bibitem[{Wang et~al.(2023a)Wang, Wu, Xie and Kozlowski}]{wang2023scalable}
\bibinfo{author}{Wang, C.}, \bibinfo{author}{Wu, X.}, \bibinfo{author}{Xie,
  Z.}, \bibinfo{author}{Kozlowski, T.}, \bibinfo{year}{2023}a.
\newblock \bibinfo{title}{Scalable inverse uncertainty quantification by
  hierarchical bayesian modeling and variational inference}.
\newblock \bibinfo{journal}{Energies} \bibinfo{volume}{16},
  \bibinfo{pages}{7664}.
\bibitem[{Wang et~al.(2023b)Wang, Fu, Du, Gao, Huang, Liu, Chandak, Liu,
  Van~Katwyk, Deac et~al.}]{wang2023scientific}
\bibinfo{author}{Wang, H.}, \bibinfo{author}{Fu, T.}, \bibinfo{author}{Du, Y.},
  \bibinfo{author}{Gao, W.}, \bibinfo{author}{Huang, K.}, \bibinfo{author}{Liu,
  Z.}, \bibinfo{author}{Chandak, P.}, \bibinfo{author}{Liu, S.},
  \bibinfo{author}{Van~Katwyk, P.}, \bibinfo{author}{Deac, A.}, et~al.,
  \bibinfo{year}{2023}b.
\newblock \bibinfo{title}{Scientific discovery in the age of artificial
  intelligence}.
\newblock \bibinfo{journal}{Nature} \bibinfo{volume}{620},
  \bibinfo{pages}{47--60}.
\bibitem[{Wicaksono(2018)}]{wicaksono2018bayesian}
\bibinfo{author}{Wicaksono, D.C.}, \bibinfo{year}{2018}.
\newblock \bibinfo{title}{Bayesian Uncertainty Quantification of Physical
  Models in Thermal-Hydraulics System Codes}.
\newblock \bibinfo{type}{Technical Report}. EPFL.
\bibitem[{Wu et~al.(2024)Wu, Chen, Zhao, Sergazinov, Li, Liu, Zhao, Xie, Guo,
  Ji et~al.}]{wu2024switchtab}
\bibinfo{author}{Wu, J.}, \bibinfo{author}{Chen, S.}, \bibinfo{author}{Zhao,
  Q.}, \bibinfo{author}{Sergazinov, R.}, \bibinfo{author}{Li, C.},
  \bibinfo{author}{Liu, S.}, \bibinfo{author}{Zhao, C.}, \bibinfo{author}{Xie,
  T.}, \bibinfo{author}{Guo, H.}, \bibinfo{author}{Ji, C.}, et~al.,
  \bibinfo{year}{2024}.
\newblock \bibinfo{title}{Switchtab: Switched autoencoders are effective
  tabular learners}.
\newblock \bibinfo{journal}{arXiv preprint arXiv:2401.02013} .
\bibitem[{Wu et~al.(2022)Wu, Tao, Zhao, Martin and
  Hovakimyan}]{wu2022optimizing}
\bibinfo{author}{Wu, J.}, \bibinfo{author}{Tao, R.}, \bibinfo{author}{Zhao,
  P.}, \bibinfo{author}{Martin, N.F.}, \bibinfo{author}{Hovakimyan, N.},
  \bibinfo{year}{2022}.
\newblock \bibinfo{title}{Optimizing nitrogen management with deep
  reinforcement learning and crop simulations}, in:
  \bibinfo{booktitle}{Proceedings of the IEEE/CVF conference on computer vision
  and pattern recognition}, pp. \bibinfo{pages}{1712--1720}.
\bibitem[{Wu et~al.(2018a)Wu, Kozlowski and Meidani}]{wu2018kriging}
\bibinfo{author}{Wu, X.}, \bibinfo{author}{Kozlowski, T.},
  \bibinfo{author}{Meidani, H.}, \bibinfo{year}{2018}a.
\newblock \bibinfo{title}{Kriging-based inverse uncertainty quantification of
  nuclear fuel performance code bison fission gas release model using time
  series measurement data}.
\newblock \bibinfo{journal}{Reliability Engineering \& System Safety}
  \bibinfo{volume}{169}, \bibinfo{pages}{422--436}.
\bibitem[{Wu et~al.(2018b)Wu, Kozlowski, Meidani and Shirvan}]{wu2018inverse}
\bibinfo{author}{Wu, X.}, \bibinfo{author}{Kozlowski, T.},
  \bibinfo{author}{Meidani, H.}, \bibinfo{author}{Shirvan, K.},
  \bibinfo{year}{2018}b.
\newblock \bibinfo{title}{Inverse uncertainty quantification using the modular
  bayesian approach based on gaussian process, part 1: Theory}.
\newblock \bibinfo{journal}{Nuclear Engineering and Design}
  \bibinfo{volume}{335}, \bibinfo{pages}{339--355}.
\bibitem[{Wu et~al.(2018c)Wu, Kozlowski, Meidani and Shirvan}]{wu2018inverse22}
\bibinfo{author}{Wu, X.}, \bibinfo{author}{Kozlowski, T.},
  \bibinfo{author}{Meidani, H.}, \bibinfo{author}{Shirvan, K.},
  \bibinfo{year}{2018}c.
\newblock \bibinfo{title}{Inverse uncertainty quantification using the modular
  bayesian approach based on gaussian process, part 2: Application to trace}.
\newblock \bibinfo{journal}{Nuclear Engineering and Design}
  \bibinfo{volume}{335}, \bibinfo{pages}{417--431}.
\bibitem[{Wu et~al.(2017a)Wu, Mui, Hu, Meidani and Kozlowski}]{wu2017inverse}
\bibinfo{author}{Wu, X.}, \bibinfo{author}{Mui, T.}, \bibinfo{author}{Hu, G.},
  \bibinfo{author}{Meidani, H.}, \bibinfo{author}{Kozlowski, T.},
  \bibinfo{year}{2017}a.
\newblock \bibinfo{title}{Inverse uncertainty quantification of trace physical
  model parameters using sparse gird stochastic collocation surrogate model}.
\newblock \bibinfo{journal}{Nuclear Engineering and Design}
  \bibinfo{volume}{319}, \bibinfo{pages}{185--200}.
\bibitem[{Wu et~al.(2017b)Wu, Wang and Kozlowski}]{mc17-2}
\bibinfo{author}{Wu, X.}, \bibinfo{author}{Wang, C.},
  \bibinfo{author}{Kozlowski, T.}, \bibinfo{year}{2017}b.
\newblock \bibinfo{title}{Global sensitivity analysis of trace physical model
  parameters based on bfbt benchmark}, in: \bibinfo{booktitle}{Proceedings of
  the MC-2017, International Conference on Mathematics Computational Methods
  Applied to Nuclear Science Engineering}, \bibinfo{address}{Jeju, Korean}.
\bibitem[{Wu et~al.(2017c)Wu, Wang and Kozlowski}]{mc17-1}
\bibinfo{author}{Wu, X.}, \bibinfo{author}{Wang, C.},
  \bibinfo{author}{Kozlowski, T.}, \bibinfo{year}{2017}c.
\newblock \bibinfo{title}{Kriging-based surrogate models for uncertainty
  quantification and sensitivity analysis}, in: \bibinfo{booktitle}{Proceedings
  of the MC-2017, International Conference on Mathematics Computational Methods
  Applied to Nuclear Science Engineering}, \bibinfo{address}{Jeju, Korean}.

\end{thebibliography}



\end{document}